\documentclass[fleqn,usenatbib]{mnras}

\usepackage{newtxtext,newtxmath}

\usepackage[T1]{fontenc}

\DeclareRobustCommand{\VAN}[3]{#2}
\let\VANthebibliography\thebibliography
\def\thebibliography{\DeclareRobustCommand{\VAN}[3]{##3}\VANthebibliography}

\usepackage{graphicx}	
\usepackage{amsmath}	
\usepackage{siunitx}
\usepackage{comment}
\usepackage{orcidlink}
\usepackage{soul}
\usepackage{anyfontsize}
\usepackage{bm}

\newcommand\Lya{Ly$\alpha$} 
\newcommand\Lyb{Ly$\beta$~}

\newcommand\xHI{$x_{\rm HI}$~}

\newcommand\HI{\hbox{H$\,\rm \scriptstyle I$}}

\newcommand\MgII{\hbox{Mg$\,\rm \scriptstyle II$}~}

\DeclareSIUnit\erg{erg}
\DeclareSIUnit\parsec{pc}
\DeclareSIUnit\propMpc{pMpc}
\DeclareSIUnit\comMpc{cMpc}
\DeclareSIUnit\angstrom{\text {Å}}

\newcommand{\MNRASreply}[1]{#1}
\newcommand{\MNRASreplyreply}[1]{#1}
\newcommand{\MNRASreplyreplyreply}[1]{#1}

\title[21-cm forest power spectrum]{Prospects of a statistical detection of the 21-cm forest and its potential to constrain the thermal state of the neutral IGM during reionization}

\author[T. Šoltinský et al.]{Tomáš Šoltinský$^{1,2}\,\orcidlink{0000-0001-7703-8929
}$\thanks{E-mail: tomas.soltinsky@inaf.it},
Girish Kulkarni$^{1}\,\orcidlink{0000-0001-5829-4716}$,
Shriharsh P. Tendulkar$^{1}\,\orcidlink{0000-0003-2548-2926}$,
James S. Bolton$^{3}\,\orcidlink{0000-0003-2764-8248}$,
\\
$^{1}$Tata Institute of Fundamental Research, Homi Bhabha Road, Mumbai 400005, India\\
$^{2}$INAF–Osservatorio Astronomico di Trieste, Via G.B. Tiepolo, 11, I-34143
Trieste, Italy\\
$^{3}$School of Physics and Astronomy, University of Nottingham, University Park, Nottingham, NG7 2RD, UK
}

\date{Accepted ---. Received ---; in original form ---}

\pubyear{2023}

\begin{document}
\label{firstpage}
\pagerange{\pageref{firstpage}--\pageref{lastpage}}
\maketitle

\begin{abstract}

\noindent
The 21-cm forest signal is a promising probe of the Epoch of Reionization complementary to other 21-cm line observables and \Lya~forest signal. Prospects of detecting it have significantly improved in the last decade thanks to the discovery of more than 30 radio-loud quasars at these redshifts, upgrades to telescope facilities, and the notion that neutral hydrogen islands persist down to $z\lesssim 5.5$.  We forward-model the 21-cm forest signal using semi-numerical simulations and incorporate various instrumental features to explore the potential of detecting the 21-cm forest at $z=6$, both directly and statistically, with the currently available (uGMRT) and forthcoming (SKA1-low) observatories. We show that it is possible to detect the 1D power spectrum of the 21-cm forest spectrum, especially at large scales of \MNRASreply{$k\lesssim8.5\,\rm MHz^{-1}$} with the $500\,\rm hr$ of the uGMRT time and \MNRASreply{$k\lesssim32.4\,\rm MHz^{-1}$} with the SKA1-low over $50\,\rm hr$ if the intergalactic medium (IGM) is $25\%$ neutral and these neutral hydrogen regions have a spin temperature of $\lesssim30\,\rm K$. On the other hand, we \MNRASreply{infer} that a null-detection of the signal with such observations of 10 radio-loud sources at $z\approx6$ can be translated into constraints on the thermal and ionization state of the IGM which are tighter than the currently available measurements. Moreover, a null-detection of the 1D 21-cm forest power spectrum with only $50\,\rm hr$ of the uGMRT observations of 10 radio-loud sources can already be competitive with the \Lya~forest and 21-cm tomographic observations in disfavouring models of significantly neutral and cold IGM at $z=6$.
\end{abstract}





\section{Introduction}


The hyperfine 21-cm line arising from the spin-flip transition in neutral hydrogen (\HI) residing in the intergalactic medium (IGM) is potentially a powerful probe of the Epoch of Reionization. This transition has a factor of $\sim10^7$ smaller cross-section than the \Lya~transition, and hence does not saturate at $z\gtrsim6$ when the IGM is expected to be at least partially neutral. This is the main advantage of the 21-cm line over the Lyman-series absorption in the spectra of luminous quasars, which is currently the premier probe of the reionization era. The 21-cm line can be observed as a series of absorption lines in spectra of distant radio-loud sources, which is known as the 21-cm forest signal. Such sources include radio-loud quasars \citep{Ciardi_2013} or gamma-ray bursts \citep{Ioka_2005,Ciardi_2015_GRB}. Given that other 21-cm line observables, such as 21-cm power spectrum from tomographic observations \citep[e.g.][]{Mertens_2020,Trott_2020,Hera_2022_obs} and sky-averaged 21-cm spectrum \citep[e.g.][]{Bowman_2018,Singh_2022}, utilize Cosmic Microwave Background as the radio background source, the 21-cm forest signal is subject to different systematic uncertainties \citep{Carilli_2002,Furlanetto_2002,Furlanetto_Oh_2006,Pritchard_2012}. Therefore, the 21-cm forest is complementary to these observables.

Not only is the 21-cm forest complementary to other probes of the Epoch of Reionization, it is also a unique probe as it is strongest in cold, neutral and slightly overdense gas \citep{Soltinsky_2021} as opposed to the \Lya~forest. \MNRASreply{This makes it one of the few probes of cold and neutral regions in the IGM besides the \MgII~line for example \citep{Hennawi_2021,Tie_2024}.} Besides the diffuse IGM, minihaloes may contribute to the 21-cm forest absorption too \citep{Meiksin_2011,Kadota_2023,Naruse_2024}, and hence in principle one can probe such small-scale structure with this signal. Furthermore, it has been suggested that the 21-cm forest signal can aid in constraining the nature of dark matter \cite[e.g.][]{Shimabukuro_2014,Shimabukuro_2020,Shimabukuro_2023,Kawasaki_2021,Shao_2023}, primordial black holes \citep{Villanueva-Domingo_2023}, thermal state of the IGM \citep[e.g.][]{Xu_2009,Xu_2011,Mack_2012,Soltinsky_2021,Shao_2023}, metal content in the IGM \citep{Bhagwat_2022} and supermassive black hole growth models \citep{Soltinsky_2023}. 

While there is no measurement of the cosmological 21-cm forest signal up to date, despite various attempts including the one by \citet{Carilli_2007}, the prospects of detecting it have improved in the last decade. Firstly, prospects of detecting the 21-cm forest have improved from the observational perspective in the form of tens of recently confirmed radio-loud quasars (RLQSO) at high redshift. Since 2020, the number of $z\geq6$ RLQSO has increased by a factor of $5$ to $\sim15$ while the number of $z\geq5.5$ RLQSO has more than tripled to $\sim34$ to date \citep{Belladitta_2020,Liu_2021,Banados_2021,Banados_2023,Banados_2024,Ighina_2021,Ighina_2023,Ighina_2024,Endsley_2023,Gloudemans_2022,Gloudemans_2023,Wolf_2024}. This number is expected to increase by orders of magnitude. For example, the physics-driven models calibrated with recent ultra-violet (UV) observations developed by \citet{Niu_2024} suggest that there are $\gtrsim50$ RLQSO with an intrinsic flux density at $150\,\rm MHz$, $S_{150}$, larger than $100\,\rm mJy$ and $\gtrsim1000$ of $S_{150}\geq10\,\rm mJy$ at $z>5.5$ over the whole sky. If large radio surveys such as The Low Frequency Array (LOFAR) Two-metre Sky Survey \citep[LoTSS][]{Shimwell_2017,Kondapally_2021}, the Tata Institute of Fundamental Research (TIFR) Giant Metrewave Radio Telescope (GMRT) Sky Survey \citep[TGSS;][]{Intema_2017}, and the Galactic and Extragalactic All-sky Murchison Widefield Array (MWA) survey \citep[GLEAM;][]{Wayth_2015} are combined with follow-up programmes such as the William Herschel Telescope Enhanced Area Velocity Explorer (WEAVE)-LOFAR survey \citep{Smith_2016_WEAVE}, these high-$z$ RLQSO should be detectable. In addition, the advancements on the instrumentational front, including the recent completion of upgraded GMRT \citep[uGMRT;][]{Gupta_2017_uGMRT} and plans for Square Kilometre Array (SKA), make the detection of the cosmological 21-cm forest even more encouraging because of their improved sensitivity.

\begin{figure*}
     \begin{minipage}{1\columnwidth}
 	  \centering
 	  \includegraphics[width=\linewidth]{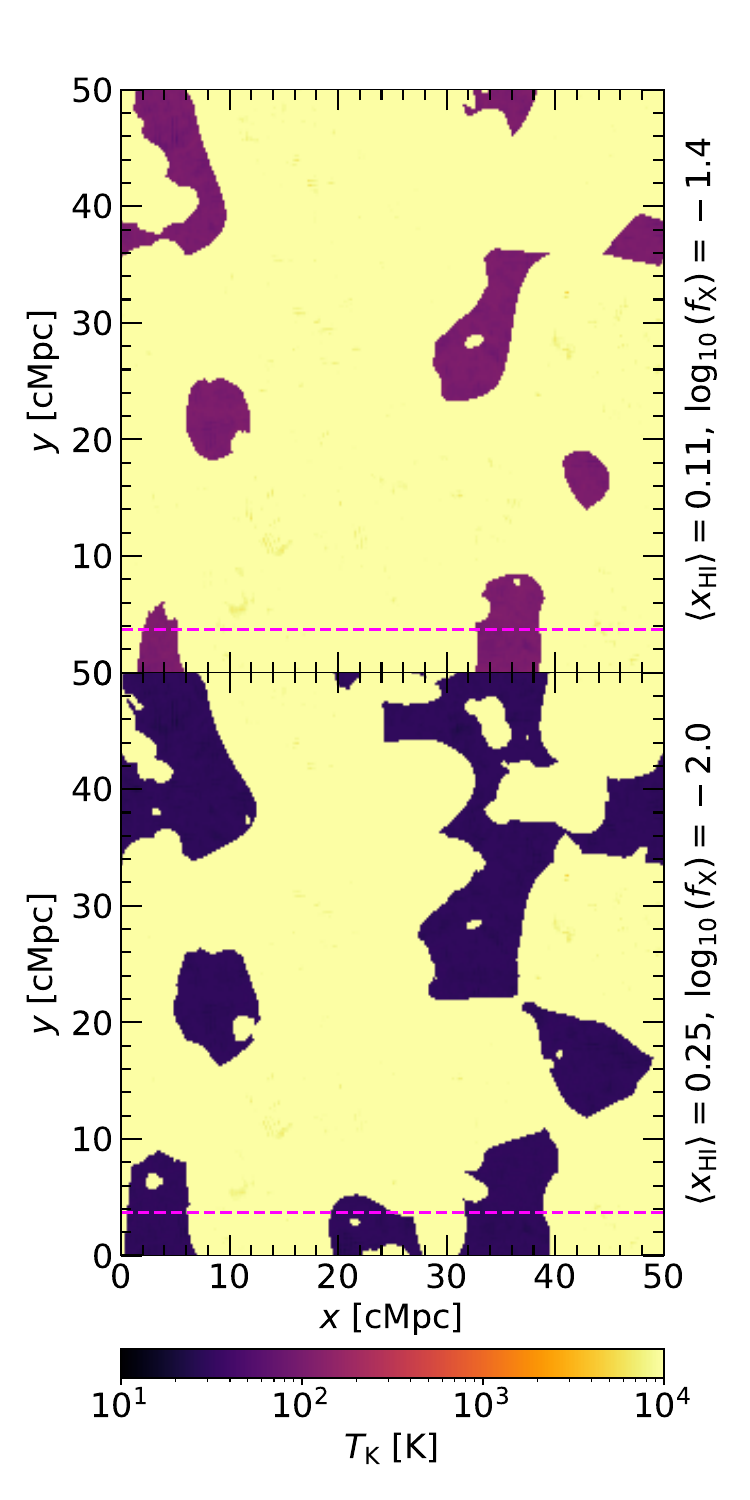}
     \end{minipage}
     \begin{minipage}{1\columnwidth}
 	  \centering
 	  \includegraphics[width=\linewidth]{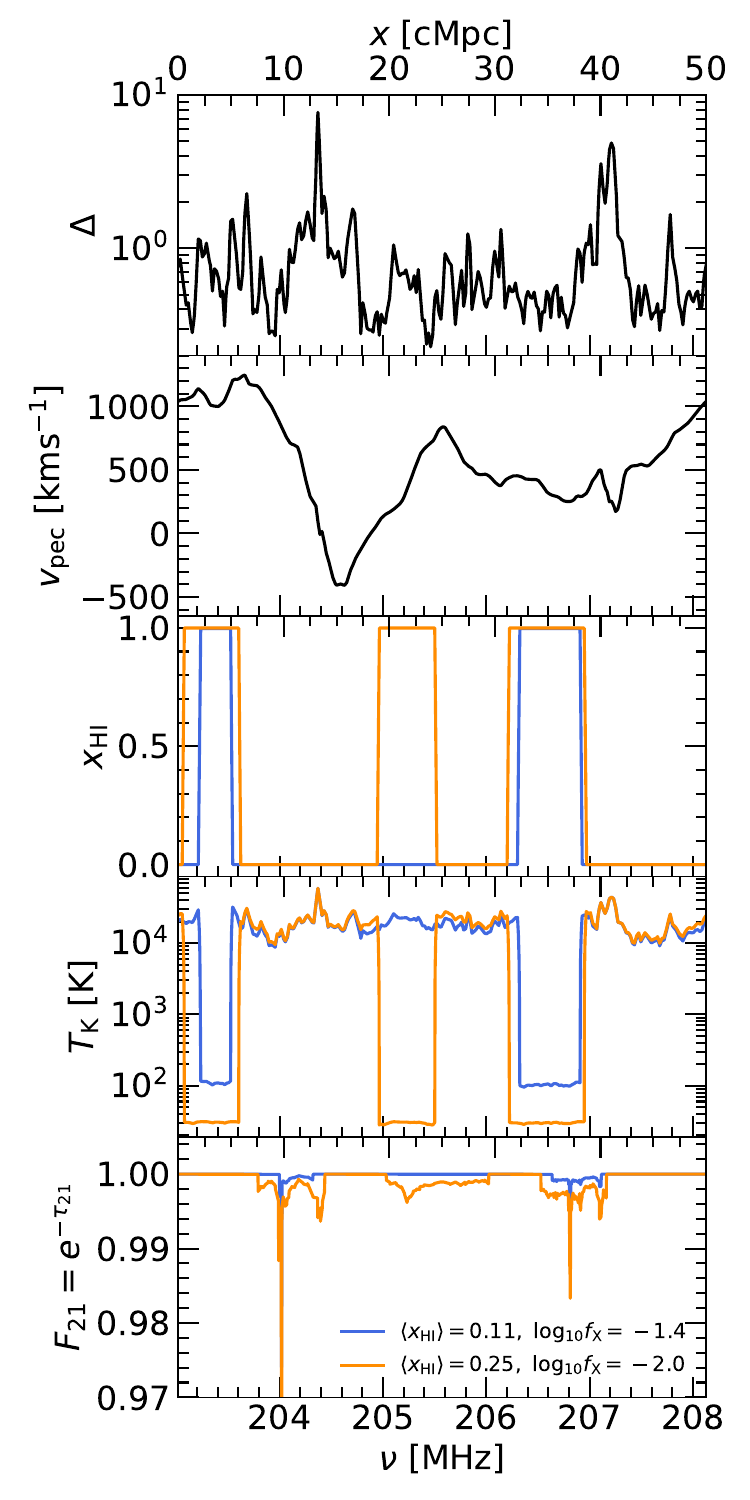}
	\end{minipage}
	\vspace{-0.3cm}
    \caption{\textit{Left}: 2D slices of the gas temperature at $z=6$ for the combination of $\langle x_{\rm HI}\rangle=0.11$ and $\mathrm{log}_{10}f_{\rm X}=-1.4$ (top panel) and $\langle x_{\rm HI}\rangle=0.25$ and $\mathrm{log}_{10}f_{\rm X}=-2.0$ (bottom panel). \textit{Right}: From top to bottom, gas overdensity, peculiar velocity, neutral hydrogen fraction, gas temperature and normalized 21-cm forest flux along the line-of-sight indicated by the dashed fuchsia lines in the left panels. The bottom three panels are shown for the cases of $\langle x_{\rm HI}\rangle=0.11$ and $\mathrm{log}_{10}f_{\rm X}=-1.4$ (\MNRASreply{blue} curves, corresponding to the top left panel) and $\langle x_{\rm HI}\rangle=0.25$ and $\mathrm{log}_{10}f_{\rm X}=-2.0$ (\MNRASreply{orange} curves, corresponding to the bottom left panel).}
    \label{fig:slice_and_LOS}
\end{figure*}

Secondly, recent \Lya~observations can be interpreted as reionization being completed by $z\leq5.6$ \citep[e.g.][]{Kulkarni_2019,Nasir_2020,Choudhury_2021,Qin_2021}. Late-end reionization picture is consistent with the large spatial fluctuations in the \Lya~forest opacity \citep{Becker_2015,Bosman2022}, deficit of \Lya~emitting galaxies around extended \Lya~absorption troughs \citep{Kashino_2020,Keating_2020,Christenson_2021}, clustering of \Lya~emitters \citep{Weinberger_2019}, thermal widths of \Lya~forest transmission spikes at $z>5$ \citep{Gaikwad_2020}, \Lya~equivalent widths \citep{Nakane_2024} and the detections of damping wings from neutral islands at $z<6$ in the \Lya~forest spectra \citep{Spina_2024,Becker_2024,Zhu_2024}. Further support of the late-end reionization models comes from the measurements of long dark gaps in the \Lya~forest \citep{Zhu_2021} and \Lyb~forest \citep{Zhu_2022}, and mean free path of ionizing photons at $z=6$ \citep{Becker_2021,Cain_2021,Zhu_2023,Gaikwad_2023}. In such late-end reionization scenario large \HI~islands are expected to persist until $z\sim5.3$. \citet{Soltinsky_2021} have shown that it might be possible to detect strong 21-cm forest absorbers at $z=6$ with SKA1-low (or even LOFAR) in the case of reionization ending at $z\simeq5.3$ and the IGM being pre-heated by X-rays such that the spin temperature $T_{\rm S}\lesssim10^2\rm\, K$. On the other hand, they suggest that a null-detection of the 21-cm forest absorption can be utilized to put a model-dependent informative lower limit on the soft X-ray background radiation. 

The goal of this study is to explore the potential of a direct and statistical detection of the 21-cm forest signal at $z\approx6$ in the light of the above-mentioned advancements in the field. While different statistics of the 21-cm forest signal have been studied \citep[e.g.][]{Mack_2012,Ewall_Wice_2014}, here we focus on the 1D power spectrum calculated from the 21-cm forest spectrum, $P_{21}$ (which has been investigated by \citet{Thyagaragan_2020}, \MNRASreply{\citet{Shao_2023,Shao_2024}} and \citet{Sun_2024}). As opposed to the previous 21-cm forest 1D power spectrum studies listed above, we focus on mock observations of (a) rather small number of background radio sources; (b) at significantly lower redshift, $z=6$ (end stages of reionization); (c) longer observational time. Furthermore, we aim to explore the constraining power of the $P_{21}$, particularly if it can be used to put joint constraints on the thermal and ionization state of the IGM. To achieve this, we take advantage of semi-numerical simulations of the reionization era IGM, \textsc{21cmfast} \citep{Mesinger_2011_21CMFAST}, to explore a wide parameter space and Bayesian statistical methods to infer parameter estimates. 

This paper is structured as follows. We begin by describing our numerical model of the IGM during the Epoch of Reionization and how we forward-model the synthetic 21-cm forest spectra in Sec.~\ref{sec:forward-model}. We then present the statistical observables of the 21-cm forest signal, particularly the differential number density of the 21-cm forest flux and 1D power spectrum in Sec.~\ref{sec:21cm_PDF} and~\ref{sec:21cm_PS}, respectively. Sec.~\ref{sec:constraining_IGM} focuses on the inference of parameters describing the thermal and ionization state of the IGM from the $P_{21}$. This study concludes with a summary in Sec.~\ref{sec:conclusions}.



\section{Forward-modelling the 21-cm forest signal}\label{sec:forward-model}


In this section we present steps how the synthetic 21-cm forest spectra are constructed in this study. In particular, we describe how the semi-numerical simulations are used to generate line-of-sight skewers of various physical quantities considering a grid of parameter combinations in Sec.~\ref{sec:IGM_sims}, how we compute the optical depths of the 21-cm forest signal in Sec.~\ref{sec:21cm_forest}, and how we forward-model realistic 21-cm forest spectra which include instrumental features of different radio telescopes in Sec.~\ref{sec:instrumental_features}.

\subsection{Modelling the intergalactic medium at \texorpdfstring{$z=6$}{}}\label{sec:IGM_sims}

To construct the synthetic 21-cm forest spectra at $z=6$ we start by running \textsc{21cmfast} semi-numerical cosmological simulations \citep{Mesinger_2011_21CMFAST,Murray_2020}. These simulations are based on the excursion set formalism and a perturbation theory approach to simulate the Epoch of Reionization universe, and hence are not computationally expensive. Therefore, this is a suitable tool for exploring a wide parameter space, which is one of the main goals of this study. The parameters of interest are the mean neutral fraction $\langle x_{\rm HI}\rangle$ and the X-ray background radiation efficiency $f_{\rm X}$. To vary the former one we fix the mean free path of ionizing photons (i.e. fix the \textsc{R\_BUBBLE\_MAX} simulation parameter) to $0.75\, \rm pMpc$ following the measurement at $z=6$ by \citet{Becker_2021} and calibrate the ionizing efficiency of high-z galaxies (i.e. varying \textsc{HII\_EFF\_FACTOR} simulation parameter) such that the $\langle x_{\rm HI}\rangle$ of interest is achieved. For the ease of comparison with literature, we implement the parametrization of the X-ray background radiation luminosity introduced in \citet{Furlanetto_2006b}
\begin{equation}
L_{\rm X}=3.4\times 10^{40}\rm\,erg\,s^{-1} \, \mathnormal{f_{\rm X}} \left(\frac{SFR}{1\,M_{\odot}\rm\,yr^{-1}}\right), \label{eq:LX} 
\end{equation}
\noindent
where the SFR is the star formation rate. In practice, we set \textsc{L\_X} simulation parameter to $\mathrm{log}_{10}\left(3.4\times 10^{40}\rm\,erg\,s^{-1}\mathnormal{f_{\rm X}}\right)$.

We generate models for combinations of 21 values of $\langle x_{\rm HI}\rangle=[0,1]$ and 26 values of $\mathrm{log}_{10}f_{\rm X}=[-4,1]$. While the $\langle x_{\rm HI}\rangle$ measurements from the \Lya~forest at $z\approx6$ including \citet{McGreer_2015}, \citet{Zhu_2022}, \citet{Gaikwad_2023}, \citet{Durovcikova_2024} and \citet{Greig_2024} are consistent with $\langle x_{\rm HI}\rangle\lesssim0.38$, our simulations cover the whole physical range of $\langle x_{\rm HI}\rangle$. The upper limits on the 21-cm power spectrum acquired by MWA, LOFAR and the Hydrogen Epoch of Reionization Array (HERA) at $z>6.5$ \citep[][respectively]{Greig_2021_MWA,Greig_2021_LOFAR,Hera_2023} disfavour scenarios of no IGM preheating. However, the X-ray background radiation is not constrained well yet, and hence we consider a large range of $\mathrm{log}_{10}f_{\rm X}$ values in this study. Also, note that X-ray photons can ionize \HI~too \citep{FurlanettoStoever2010} which results in $\langle x_{\rm HI}\rangle$ significantly lower than 1 for high $\mathrm{log}_{10}f_{\rm X}$ even if \MNRASreply{the \textsc{HII\_EFF\_FACTOR}} of 0 is implemented. For example, if $\mathrm{log}_{10}f_{\rm X}=1$ at $z=6$ in our models, we obtain $\langle x_{\rm HI}\rangle\lesssim0.82$, and to achieve $\langle x_{\rm HI}\rangle>0.99$ in our models $\mathrm{log}_{10}f_{\rm X}\leq-0.4$ is required. \MNRASreply{This results in the fact that some of the high $\langle x_{\rm HI}\rangle$ and high $\mathrm{log}_{10}f_{\rm X}$ models are not acquired. In summary, we generate 534 models in total.}

These simulations are run on a grid of $256^3$ pixels with a volume of \MNRASreply{$\left(50\, \rm cMpc\right)^3$} at $z=6$. Note that these simulations do not capture coherent regions of \HI~islands on scales larger than the simulation box. The simulations are based on a flat $\Lambda$CDM cosmology consistent with the measurements of \citet{planck2014}, namely $\Omega_{\Lambda}=0.692$, $\Omega_{\rm m}=0.308$, $\Omega_{\rm b}=0.0482$, $\sigma_8=0.829$, $n_{\rm s} =0.961$ and $h=0.678$. Furthermore, a primordial helium fraction by mass of $Y_{\rm p}=0.24$ \citep{Hsyu_2020} is adopted. Two examples of 2D slices of the gas kinetic temperature, are presented in the left panels of Fig.~\ref{fig:slice_and_LOS}. The top panel corresponds to a warmer and more ionized IGM ($\mathrm{log}_{10}f_{\rm X}=-1.4$, $\langle x_{\rm HI}\rangle=0.11$) than the bottom one ($\mathrm{log}_{10}f_{\rm X}=-2$, $\langle x_{\rm HI}\rangle=0.25$). In both cases the ionized regions (yellow areas) have a temperature higher than $2\times10^4\,\rm K$, and hence the 21-cm forest is expected to be suppressed completely. More importantly, in the top panel the mean temperature of neutral regions $T_{\rm HI}=102\,\rm K$\footnote{The \textsc{21cmfast} simulates reionization bubbles using excursion set principle, and hence the IGM can have neutral hydrogen fraction of $x_{\rm HI}=0$ (completely ionized bubbles) and $x_{\rm HI}\approx1$ (fully neutral islands) and no values between. Therefore, there is no arbitrary criterion for what is defined as a neutral region.} and in the bottom panel $T_{\rm HI}=30\,\rm K$. We extract $1000$ periodic lines-of-sight parallel to the simulation box which contain overdensity $\Delta=1+\delta$, peculiar velocity $v_{\rm pec}$, neutral hydrogen fraction \xHI and $T_{\rm K}$ fields. The right panels of Fig.~\ref{fig:slice_and_LOS} show these quantities from top to bottom, respectively, for the LOS piercing the IGM as indicated by the dashed fuchsia line in the left panels.

\subsection{The 21-cm forest absorption spectrum}\label{sec:21cm_forest}

The ground state of neutral hydrogen atoms is split into two energy levels given by the relative orientation of the proton and electron spin. The energy difference corresponds to the energy of a photon with a rest-frame wavelength of $\lambda_{21}=21.11\,\rm cm$ or equivalently a frequency of $\nu_{21}=1420.41\,\rm MHz$. This hyperfine structure of \HI~atoms gives rise to the 21-cm line. Here we focus on the 21-cm forest signal. Following \citet{Soltinsky_2021}, we compute the normalized 21-cm forest flux, $F_{21}=\rm e^{-\tau_{21}}$, where $\tau_{21}$ is the optical depth to the 21-cm photons. In the discrete form, the $\tau_{21}$ for pixel $i$ is given by \citep[e.g.][]{Furlanetto_2002}
\begin{align}
\tau_{\rm 21, i} =~& \frac{3h_{\rm p}c^{3}A_{10} }{32\pi^{3/2}\nu_{21}^{2}k_{\rm B}} \frac{\delta v}{H(z)} \nonumber \\
             & \times \sum_{j=1}^{N}\frac{n_{{\rm HI}, j}}{b_{j}T_{{\rm S}, j}}\exp\left( - \frac{ (v_{{\rm H},i}-u_{j})^{2}} {b^{2}_{j}}\right), \label{eq:tau21_discrete}
\end{align}
\noindent
where $h_{\rm p}$ and $k_{\rm B}$ are the Planck and Boltzmann constants, respectively, $c$ is speed of light, $A_{10}=2.85\times 10^{-15}\rm\,s^{-1}$ is the Einstein spontaneous emission coefficient for the hyperfine transition, $\delta v$ is the velocity width of the pixels\footnote{We resample the LOS of IGM fields using linear interpolation to ensure that the 21-cm absorption line profiles are converged correctly. For more details we refer the reader to \citet{Soltinsky_2021}.}, $H(z)$ is the Hubble parameter, $n_{\rm HI}$ is the \HI~number density and $b=(2k_{\rm B}T_{\rm K}/m_{\rm H})^{1/2}$ is the Doppler parameter. We assume fully coupled spin temperature, $T_{\rm S}$, to $T_{\rm K}$ (i.e. $T_{\rm S}=T_{\rm K}$). This is a valid assumption at $z=6$ because \Lya~background is expected to be strong at such low redshift. Note that \citet{Soltinsky_2021} found differences in the 21-cm forest signal between the case of $T_{\rm S}=T_{\rm K}$ and $T_{\rm S}\neq T_{\rm K}$, however at $z=6$ they were rather small. In Eq.~\ref{eq:tau21_discrete} we convolve over the Gaussian line profile to include not only the effect of Hubble velocity, $v_{\rm H}$, on the 21-cm forest, but also the effect of peculiar velocity of the gas along the LOS, in $u_{j}=v_{{\rm H},j}+v_{{\rm pec},j}$. In the bottom right panel of Fig.~\ref{fig:slice_and_LOS} we show an example synthetic 21-cm forest spectrum at $z=6$ (without any instrumental features) for the case of $\mathrm{log}_{10}f_{\rm X}=-2$ and $\langle x_{\rm HI}\rangle=0.25$ (orange curve) and $\mathrm{log}_{10}f_{\rm X}=-1$ and $\langle x_{\rm HI}\rangle=0.11$ (blue curve). It is clearly seen how the ionization bubbles completely eliminate the 21-cm forest absorbers while the higher spin temperature suppresses them.


\subsection{Instrumental features of the uGMRT and SKA1-low} \label{sec:instrumental_features}

\begin{figure}
     \begin{minipage}{1\columnwidth}
 	  \centering
 	  \includegraphics[width=\linewidth]{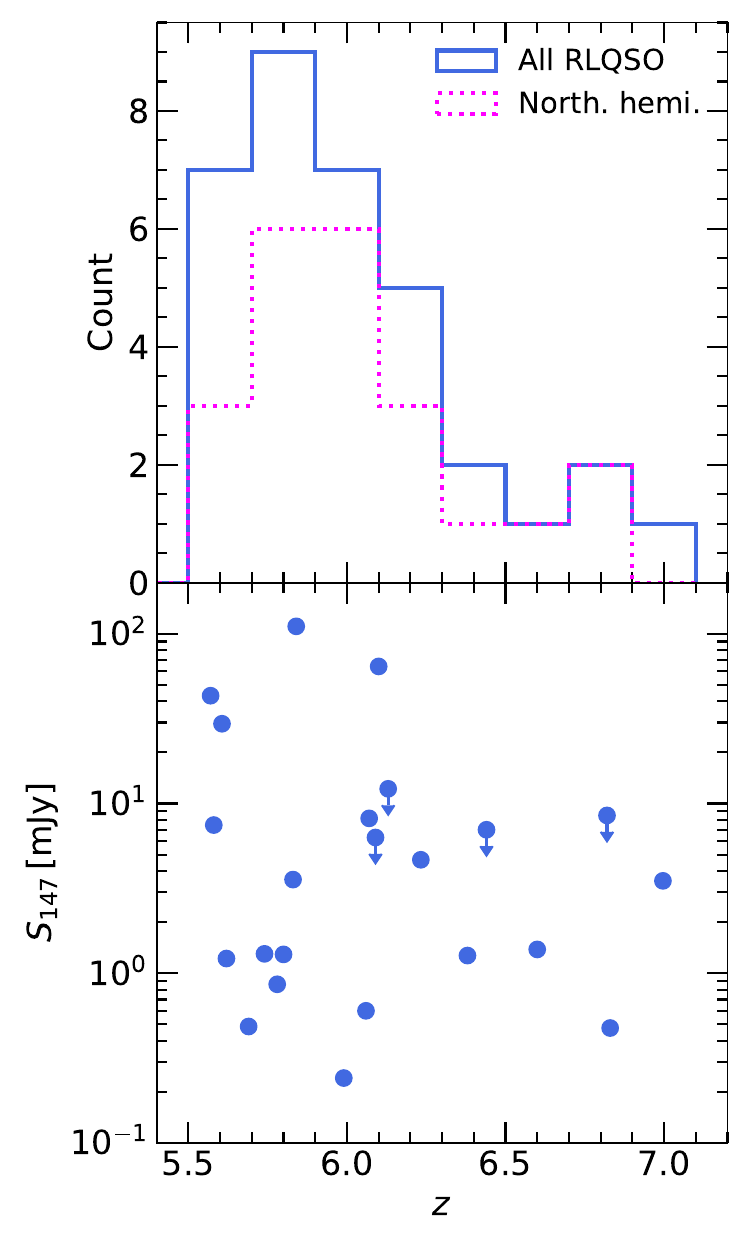}
	\end{minipage}
	\vspace{-0.3cm}
    \caption{\textit{Top panel}: Redshift distribution of radio-loud quasars identified at $z\geq5.5$ across the whole sky (solid blue curve) and in the northern hemisphere (dashed fuchsia curve). The sample is compiled from \citet{Fan_2001}, \citet{McGreer_2006}, \citet{Willott_2010}, \citet{Zeimann_2011}, \citet{Banados_2015,Banados_2018_z584QSO,Banados_2021,Banados_2023,Banados_2024}, \citet{Belladitta_2020}, \citet{Liu_2021}, \citet{Ighina_2021,Ighina_2023,Ighina_2024}, \citet{Endsley_2023}, \citet{Gloudemans_2022,Gloudemans_2023} and \citet{Wolf_2024}. \textit{Bottom panel}: Intrinsic flux density at $147\,\rm MHz$ of these radio-loud quasars as a function of their redshift. The arrows indicate quasars for which only upper limits on $S_{147}$ are available. Note that not all quasars have a $S_{147}$ measurement.}
    \label{fig:RLQdist}
\end{figure}

Before we describe how we incorporate instrumental features in our forward-modelled 21-cm forest spectra, we note that there is a substantial number of radio-loud sources at high $z$, particularly $34$ RLQSOs at $z\geq5.5$ to date\footnote{The whole sample can be found at \href{https://tomassoltinsky.github.io//}{https://tomassoltinsky.github.io}.}. The redshift distribution of these RLQSOs is shown in Fig.~\ref{fig:RLQdist} with the highest redshift reaching $z=6.82$ \citep{Banados_2021}. Most of these RLQSOs are located in the northern hemisphere as indicated by the dashed fuchsia curve. Given that this sample lies within the the redshift range of $5.5\lesssim z\lesssim7$, we will consider the observed frequencies of $177.5\,\mathrm{MHz}\lesssim \nu_{\rm obs}\lesssim220\,\mathrm{MHz}$. 

\begin{figure}
     \begin{minipage}{1\columnwidth}
 	  \centering
 	  \includegraphics[width=\linewidth]{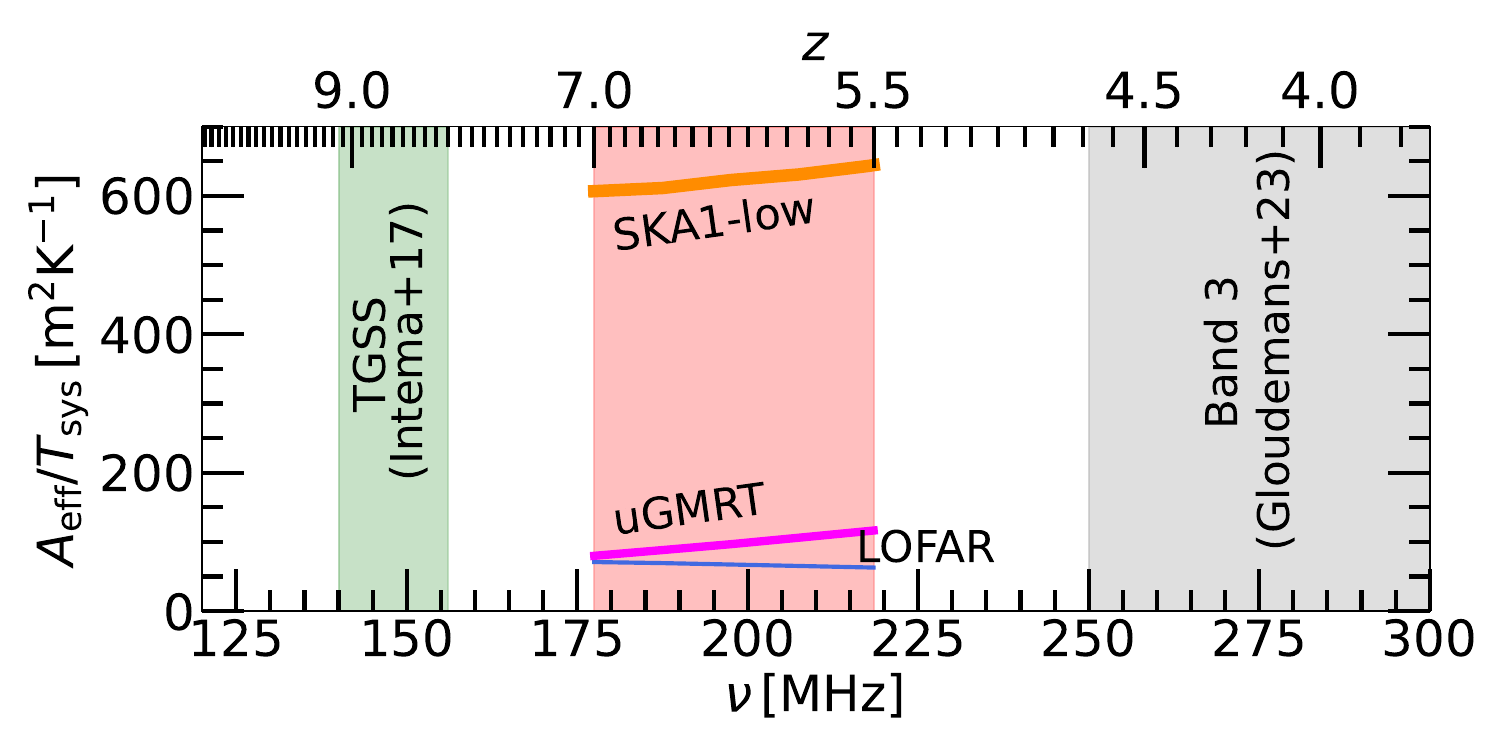}
	\end{minipage}
	\vspace{-0.3cm}
    \caption{The sensitivity in terms of $A_{\rm eff}/T_{\rm sys}$ \MNRASreplyreply{for the whole array} at different observed frequencies for the LOFAR (thin blue curve), uGMRT (semi-thick fuchsia curve) and SKA1-low (thick orange curve) taken from \citet{Braun_2019}. The sensitivities of these three telescopes are shown only in the frequency range in which the 21-cm forest is expected to be currently observable, i.e. at $177.5\,\rm MHz\lesssim \nu_{\rm obs}\lesssim218.5\,\rm MHz$ (red shaded region). These frequencies correspond to the redshifts of the quasar sample shown in Fig.~\ref{fig:RLQdist}. The green shaded region indicates the frequencies at which the TGSS \citep{Intema_2017} was performed and the grey shaded region partially marks the band-3 of the uGMRT which was used for the survey in \citet{Gloudemans_2023}.}
    \label{fig:telescope_sensitivity}
\end{figure}

In Fig.~\ref{fig:telescope_sensitivity} we show the sensitivity of various telescopes \MNRASreplyreply{assuming the whole array}, $A_{\rm eff}/T_{\rm sys}$, as a function of observed frequency with the above frequency range indicated with the red shaded region. These values are taken from \citet{Braun_2019} (see their fig. 8) and are shown for LOFAR (blue curve), band-2 of uGMRT (fuchsia curve) and SKA1-low (orange curve). For comparison, the green and grey shaded regions mark some of the past GMRT surveys, particularly the TGSS by \citet{Intema_2017} and the one by \citet{Gloudemans_2023} in the uGMRT band-3, respectively. It is clear that at these frequencies the uGMRT is currently the most sensitive operational telescope. Therefore, we focus our analyses on mock observations by uGMRT in band-2 to explore the potential of the currently achievable observations of the 21-cm forest signal. However, we extend our analyses to the SKA1-low to probe future outlooks too.

\begin{figure*}
    \begin{minipage}{1\columnwidth}
 	  \centering
 	  \includegraphics[width=\linewidth]{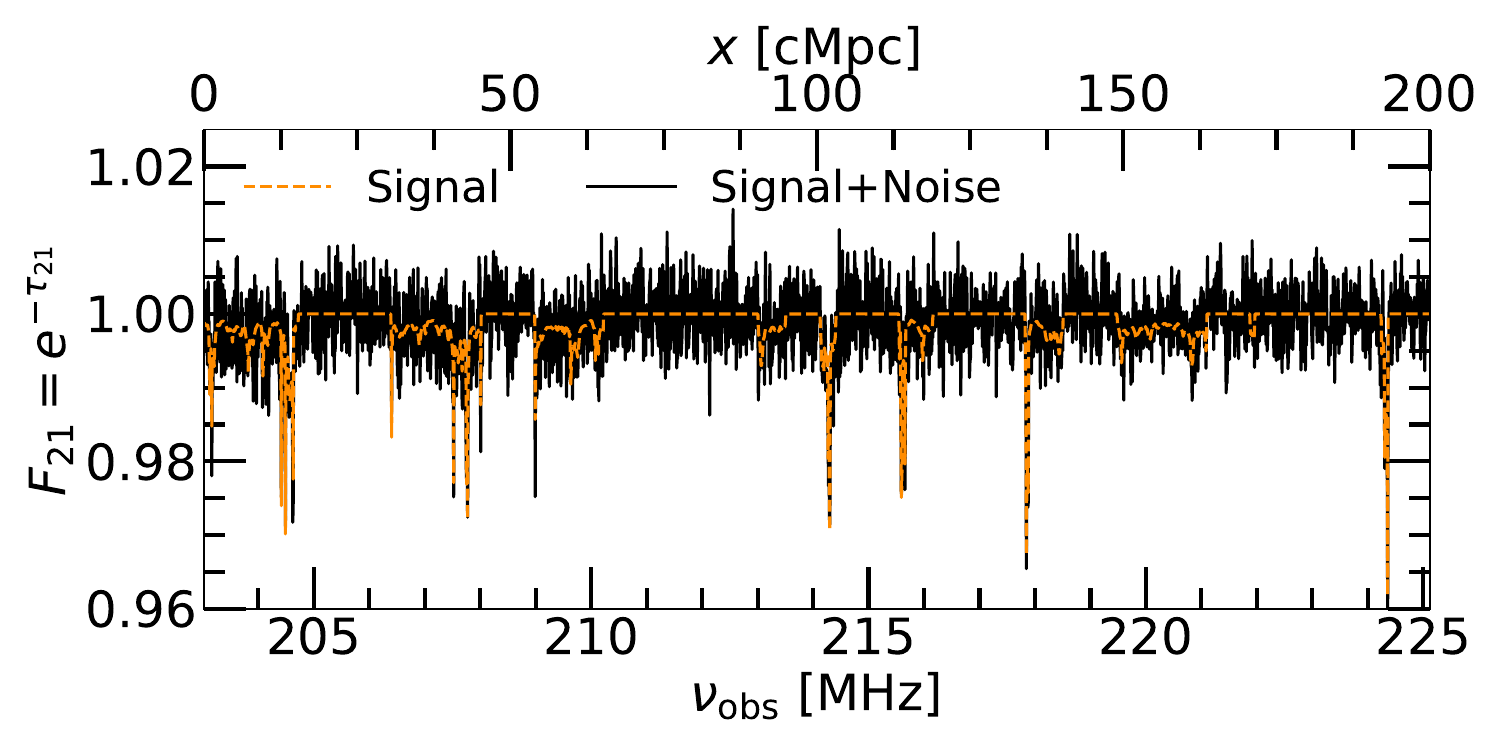}
	\end{minipage}
     \begin{minipage}{1\columnwidth}
 	  \centering
 	  \includegraphics[width=\linewidth]{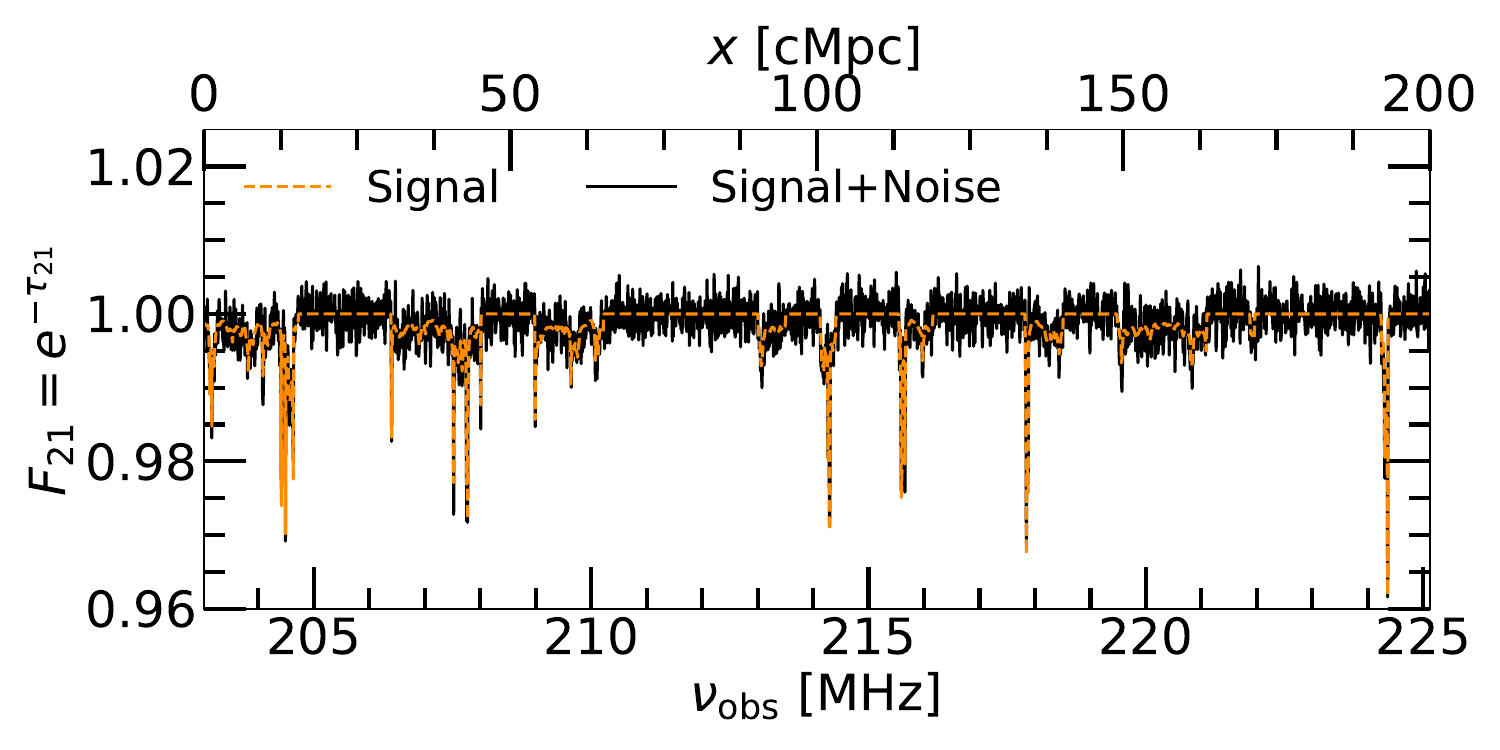}
	\end{minipage}
	\vspace{-0.3cm}
    \caption{The forward-modelled 21-cm forest signal (\MNRASreply{solid  black curves}) in the spectrum of quasars with $S_{147}=64.2\,\rm mJy$ and $\alpha_{\rm R}=-0.44$ at $z=6$ as observed by the uGMRT over $t_{\rm int}=500\,\rm hr$ (left panel) and SKA1-low over $t_{\rm int}=50\,\rm hr$ (right panel). We also show the simulated physical signal without added noise in dashed orange curves. In both the cases, noisy spectrum and spectrum without the telescope \MNRASreply{noise}, the spectrum has been smoothed by a boxcar function of width $\Delta\nu=8\,\rm kHz$ to model the spectral resolution of the telescopes. \MNRASreply{In all cases a model of $\mathrm{log}_{10}f_{\rm X}=-2$ and $\langle x_{\rm HI}\rangle=0.25$ has been used.}}
    \label{fig:noisyspectrum}
\end{figure*}

Firstly, we model the effect of spectral resolution of the telescope on the observed 21-cm forest spectrum by convolving the spectrum with a boxcar function of $\Delta\nu$ width. The band-2 of uGMRT covers the frequency range of $125-250\,\rm MHz$ and contains 16384 spectral channels \citep{Gupta_2017_uGMRT}. Splitting the whole bandwidth by the number of spectral channels, one can achieve the spectral resolution of $\Delta\nu=8\,\rm kHz$, which we will use in our analyses. A smoothed spectrum for $\mathrm{log}_{10}f_{\rm X}=-2$ and $\langle x_{\rm HI}\rangle=0.25$ is shown as the dashed orange curve in Fig.~\ref{fig:noisyspectrum}. Comparing this with unsmoothed spectrum in Fig.~\ref{fig:slice_and_LOS}, the main effect of smoothing is the suppression of the 21-cm forest absorption features depth. Note that we constructed longer spectra than in Fig.~\ref{fig:slice_and_LOS} by splicing 4 randomly chosen $50\,\rm cMpc$ long spectra. This way we forward model spectra of $200\,\rm cMpc$ and $22.1\,\rm MHz$ bandwidth at $z=6$ which will be important in Section~\ref{sec:21cm_PS} and~\ref{sec:constraining_IGM}.

Secondly, we incorporate the telescope noise. In particular, we add a Gaussian white noise with the rms given by \citep[cf.][]{Datta_2007,Ciardi_2013}
\MNRASreplyreply{\begin{equation}
    \sigma_{\rm N} = \left(\frac{A_{\rm eff}}{T_{\rm sys}}\right)_{N_{\rm d}=1}^{-1}\frac{\sqrt{2}k_B}{\sqrt{N_{\rm d}(N_{\rm d}-1)\Delta\nu t_{\rm int}}}, \label{eq:noise}
\end{equation}}
\noindent
where \MNRASreplyreply{$\left(A_{\rm eff}/T_{\rm sys}\right)_{N_{\rm d}=1}$ is the frequency-dependent sensitivity for a single dish} taken from \citet{Braun_2019}\footnote{While we include the dependence of $A_{\rm eff}/T_{\rm sys}$ on the $\nu_{\rm obs}$, the $A_{\rm eff}/T_{\rm sys}$ does not vary significantly over our synthetic spectra given that their frequency range is narrow. Note also that \citet{Braun_2019} gives the $A_{\rm eff}/T_{\rm sys}$ for the whole array of dishes instead of a single dish. \MNRASreplyreply{Therefore, we define $\left(A_{\rm eff}/T_{\rm sys}\right)_{N_{\rm d}=1}=A_{\rm eff}/T_{\rm sys}\sqrt{N_{\rm d,tot}(N_{\rm d,tot}-1)}$, where $N_{\rm d,tot}$ is the number of dishes of the whole array of the observatory, i.e. $30$ and $512$ for the uGMRT and SKA1-low, respectively.}}, $N_{\rm d}$ is the number of dishes and $t_{\rm int}$ is the integration time\footnote{Instead, one can use GMRT Exposure Time Calculator (ETC, \href{http://www.ncra.tifr.res.in:8081/~secr-ops/etc/rms/rms.html}{http://www.ncra.tifr.res.in:8081/secr-ops/etc/rms/rms.html}) to compute the $\sigma_{\rm N}$. We have compared our calculation with the GMRT ETC and found that we overestimate the $\sigma_{\rm N}$ by $22\%$. \MNRASreply{There is a sensitivity calculator for the SKA1-low too (\href{https://sensitivity-calculator.skao.int/low}{https://sensitivity-calculator.skao.int/low}) which results in $\sim30\%$ larger $\sigma_{\rm N}$ than our calculation.}}.

\begin{figure*}
    \begin{minipage}{0.33\linewidth}
 	  \centering
 	  \includegraphics[width=\linewidth]{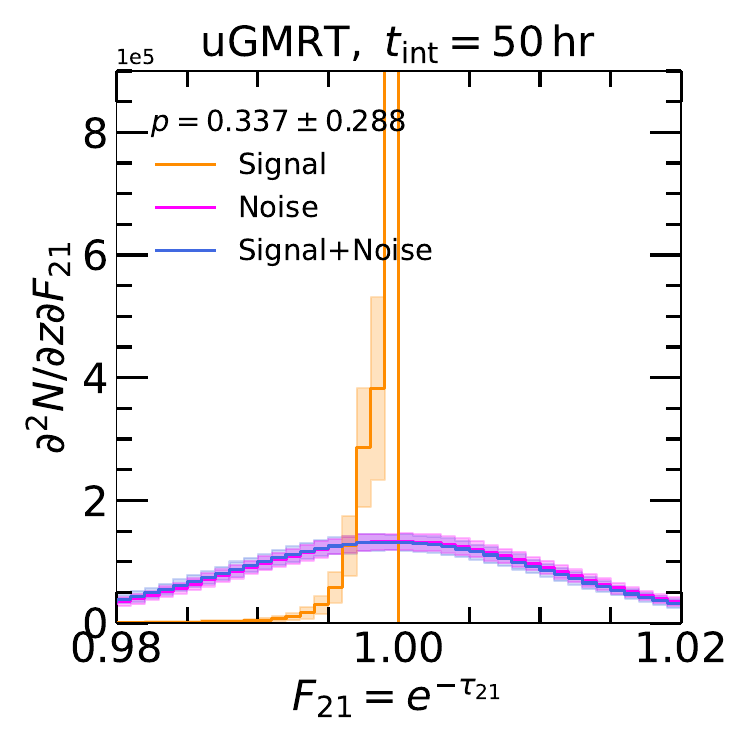}
	\end{minipage}
    \begin{minipage}{0.33\linewidth}
 	  \centering
 	  \includegraphics[width=\linewidth]{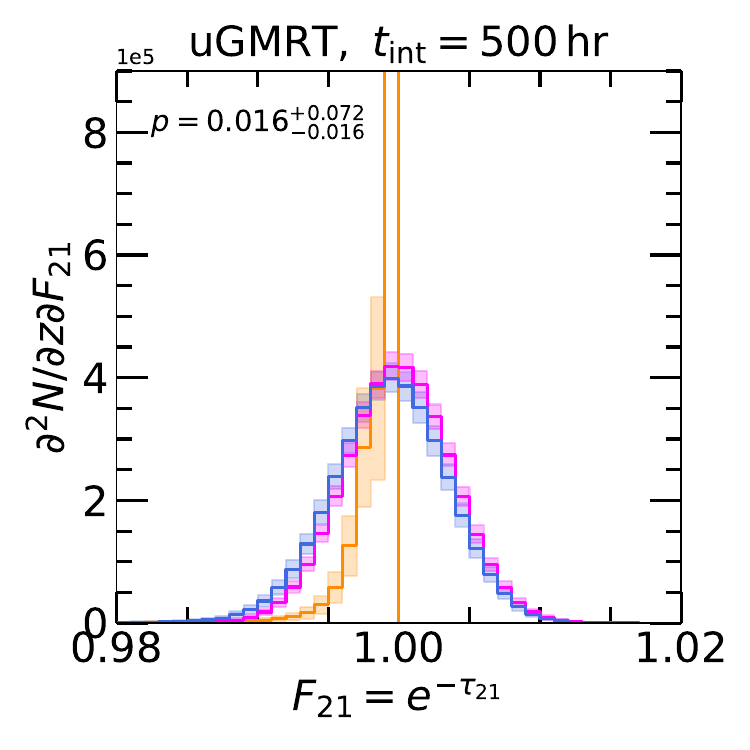}
	\end{minipage}
    \begin{minipage}{0.33\linewidth}
 	  \centering
 	  \includegraphics[width=\linewidth]{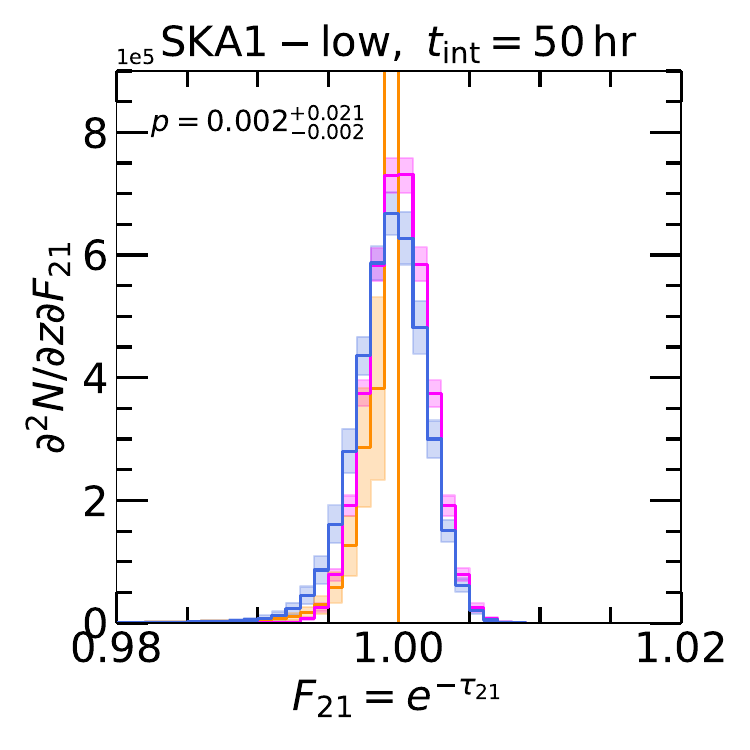}
	\end{minipage}
	\vspace{-0.3cm}
    \caption{Differential number density distribution of 21-cm forest transmission by frequency channel for a spectrum containing purely the signal (orange curves), purely the noise (fuchsia curves) and
    an observed spectrum (signal and noise, blue curves). The solid lines correspond to the median values while the shaded regions indicate the 68 per cent range computed from 1000 mock spectra. Spectra were modelled for an observation of a background source with $S_{147}=64.2\,\rm mJy$ and $\alpha_{\rm R}=-0.44$ by the uGMRT over $t_{\rm int}=50\,\rm hr$ (left panel) and $500\,\rm hr$ (middle panel) and by the SKA1-low over $50\,\rm hr$ (right panel). A two-sided Kolmogorov-Smirnov test $p$-value for the observed and noise distribution is shown too.}
    \label{fig:PDF_multiLOS}
\end{figure*}

The 21-cm forest spectrum is then normalized based on the intrinsic spectrum of the background radio source. We model the radio spectrum of the RLQSO as a single power-law with index $\alpha_{\rm R}$ and normalization given by the intrinsic flux density at $147\,\rm MHz$, $S_{147}$, i.e. $S=S_{147}\left(\nu/147\,\mathrm{MHz}\right)^{\alpha_{\rm R}}$. Note that this will make the normalized rms of the telescope noise frequency-dependent. Both $S_{147}$ and $\alpha_{\rm R}$ have been measured for various high-$z$ RLQSO by \citet{Gloudemans_2022,Gloudemans_2023} and \citet{Banados_2023} to name few examples. \MNRASreply{The measurements of the $S_{147}$} are provided in the bottom panel of Fig.~\ref{fig:RLQdist}. Note that most of the $z>5.5$ RLQSOs are rather faint ($S_{147}\sim1\,\rm mJy$), however there are 4 with $S_{147}>29\,\rm mJy$. In what follows we will consider the PSO J0309+27 blazar, one of the brightest confirmed radio-loud sources at $z\approx6$, which was measured to have $S_{147}=64.2\pm6.2\,\rm mJy$ and $\alpha_{\rm R}=-0.44\pm0.11$ \citep{Belladitta_2020}.

The left panel of Fig.~\ref{fig:noisyspectrum} shows a forward-modelled 21-cm forest spectrum implemented with the above described instrumental features assuming an $t_{\rm int}=500\,\rm hr$ observation by the uGMRT (solid black curve) \MNRASreply{and the same LOS as for the case of no telescope noise (dashed orange curve, IGM model of $\mathrm{log}_{10}f_{\rm X}=-2$ and $\langle x_{\rm HI}\rangle=0.25$)}. We assumed the whole array of $N_{\rm d}=30$. \MNRASreply{The signal-to-noise ratio (SNR) of the deviation from the continuum (i.e. absorption signal) reaches $11.9$ at $\nu\approx224.4\,\rm MHz$ while the highest SNR for all 1000 synthetic LOS is $36.5$. However, frequency channels with such high SNR are $>20$ times less abundant than the ones at more representative values of $\mathrm{SNR}\approx0.4$. More details about the SNR calculation can be found in Appendix~\ref{app:SNR}. Given this, it is obvious that the direct detection of individual absorption features is challenging.} In addition, we do not consider the effect of the Radio Frequency Interference (RFI). RFI can introduce both narrow-band and broad-band features in the radio spectra which would complicate observations of the 21-cm forest signal. \MNRASreply{However, according to \textit{The Exposure Time Calculator for the upgraded Giant Metrewave Radio Telescope} handbook\footnote{\href{http://www.ncra.tifr.res.in:8081/~secr-ops/etc/etc_help.pdf}{http://www.ncra.tifr.res.in:8081/$\sim$secr-ops/etc/etc\_help.pdf}} the frequencies unusable due to the RFI span over $165\,\rm MHz-190\,\rm MHz$ which are lower than the frequencies of interest as seen in Fig.~\ref{fig:telescope_sensitivity}.}


We now assume the same \MNRASreply{LOS and instrumental} behaviour for the SKA1-low observatory (i.e. using Eq.~\ref{eq:noise}) and the whole interferometer with $N_{\rm d}=512$ but shorter observational time of $t_{\rm int}=50\,\rm hr$. The $A_{\rm eff}/T_{\rm sys}$ for the SKA1-low is taken from \citet{Braun_2019} too. Given that the SKA1-low is expected to observe sky south from Dec. +30° \citep{Zheng_2020}, the PSO J0309+27 should be observable by this telescope too. \MNRASreply{The observations at $72\,\rm MHz-231\,\rm MHz$ by the MWA, which is located at the same site as the SKA1-low is planned to be built, show that this area is significantly less impacted by the RFI than LOFAR and uGMRT sites \citep{Offringa_2015}.} Besides this, we see in the right panel of Fig.~\ref{fig:noisyspectrum} that the noise is not as severe even if the observational time is 10 times shorter than assumed for the uGMRT observation. \MNRASreply{However, while the highest SNR reaches $19.0$ in this particular LOS and $63.3$ in the whole sample of simulated LOS, similarly to the previous case, such frequency channels are $>20$ times rarer than the bulk of the channels which has $\mathrm{SNR}\approx0.9$.}



\section{The 21-cm forest statistics}\label{sec:21cm_stats}


Motivated by the apparent difficulty of the direct detection of individual 21-cm forest absorption features in the spectra of even one of the brightest high-$z$ radio-loud source we currently know we aim to investigate the prospects of a statistical detection of the 21-cm forest signal. In particular, we consider the differential number density of $F_{21}$ in Sec.~\ref{sec:21cm_PDF} and 1D power spectrum from the 21-cm forest spectrum in Sec.~\ref{sec:21cm_PS}.

\subsection{Differential number density distribution of the 21-cm forest transmission}\label{sec:21cm_PDF}

Given that the background radio-source has a high brightness temperature, we expect the 21-cm forest signal always in absorption. Therefore, the frequency channel (or pixel) differential number density of the $F_{21}$ of the spectrum containing purely just the signal will be at $F_{21}\leq1$. This is shown in Fig.~\ref{fig:PDF_multiLOS} where we show differential distributions of $F_{21}$ by pixel of the 21-cm forest signal spectrum without (orange curves) and with the telescope noise included (blue curves) and telescope noise only (without the signal, fuchsia curves). The solid curves are mean values over 1000 LOS and the shaded regions correspond to the 68 percent scatter around the mean. Note that this is different from the distributions shown in studies like \citet{Shimabukuro_2014} and \citet{Soltinsky_2021} as they considered distribution of individual absorption features and we are interested in the distribution of the $F_{21}$ at each frequency channel. One can see that the signal only distribution (orange curves) extends only to $F_{21}\leq1$ as expected.

On the other hand, the telescope noise is modelled as a Gaussian white noise as described in Section~\ref{sec:instrumental_features}. Hence, it is distributed as a normal distribution function with a mean at $F_{21}=1$ in both absorption and emission (higher and lower than $F_{21}=1$). If a measurement of a radio spectrum at relevant frequencies is obtained, one would expect the distribution of the real $F_{21}$ (signal and noise, blue curves) to deviate from the noise only distribution at $F_{21}<1$ if such spectrum contained the 21-cm forest signal. Therefore, a deviation from the normal distribution can be considered to be a statistical detection of the 21-cm forest signal.

To test if the observed spectrum (signal and noise) distribution is consistent with the noise distribution, we use a two-sided Kolmogorov-Smirnov test. If the Kolmogorov-Smirnov test $p$-value is more than 0.05, it is likely that the observed spectrum distribution comes from the same distribution as the noise $F_{21}$, and therefore this would be considered a non-detection. An example of this is shown in the left panel of Fig.~\ref{fig:PDF_multiLOS}, where we considered an observation by the uGMRT over $50\,\rm hr$. The mean $p=0.337$ and $\approx85\%$ of the 1000 synthetic spectra have $p>0.05$. We \MNRASreply{infer} that with such observational setup it is unlikely to achieve a detection this way. However, if the uGMRT observations are extended 10-fold in time (i.e. $t_{\rm int}=500\,\rm hr$), less than $10\%$ of the 21-cm forest spectra results in a non-detection (the middle panel). This is further decreased to only few per cents of the spectra in the case of $t_{\rm int}=50\,\rm hr$ with the SKA1-low. Not only that, but in the right panel of Fig.~\ref{fig:PDF_multiLOS} it is clearly visible how the signal+noise curve deviates from the noise only distribution.

\subsection{The 21-cm forest power spectrum}\label{sec:21cm_PS}

\begin{figure}
    \begin{minipage}{\linewidth}
 	  \includegraphics[width=\linewidth]{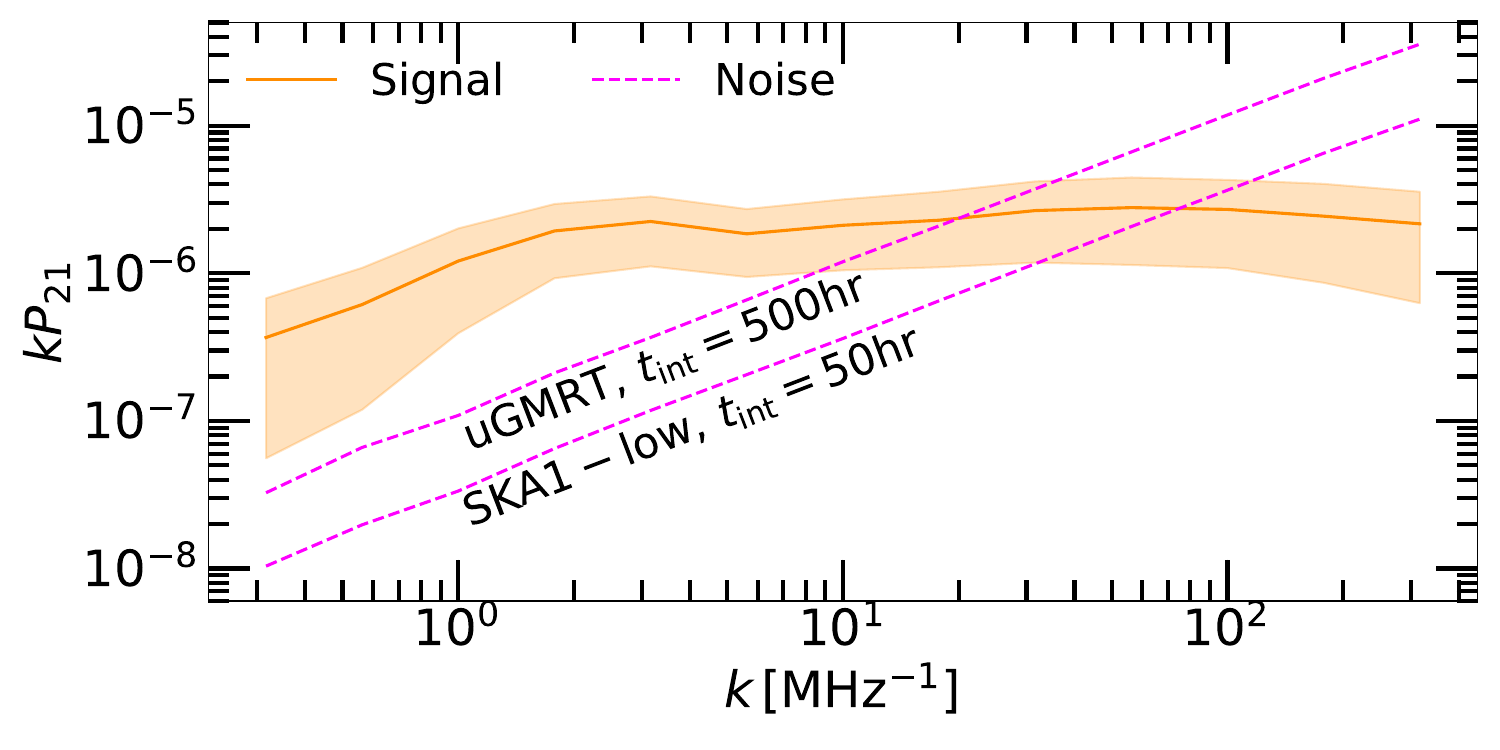}
	\end{minipage}
	\vspace{-0.3cm}
    \caption{1D power spectrum from a single $50\,\rm cMpc$ long 21-cm forest spectrum in the fiducial model of the IGM with $\mathrm{log}_{10}f_{\rm X}=-2$ and $\langle x_{\rm HI}\rangle=0.25$. The solid orange curve shows the median power spectrum of the signal and the orange shaded region marks 68 per cent range from 1000 mock spectra. The dashed fuchsia curve shows the power spectrum of the noise assuming an observation targeting $S_{147}=64.2\,\rm mJy$ and $\alpha_{\rm R}=-0.44$ quasar with uGMRT over $t_{\rm int}=500\,\rm hr$ (top curve) and SKA1-low over $t_{\rm int}=50\,\rm hr$.}
    \label{fig:1DPS_noise}
\end{figure}

In this section we turn to the computation of the 1D power spectrum of the transmitted flux using the estimator $\delta_{\rm F}=F_{21}-1=\rm e^{-\tau_{21}}-1$. This 1D 21-cm forest power spectrum, $P_{21}$, is defined as
\begin{equation}
    P_{21}\left(k\right)\delta_{\rm D}\left(k-k'\right)=\left\langle\tilde{\delta}_{\rm F}\left(k\right)\tilde{\delta}_{\rm F}^*\left(k'\right)\right\rangle, \label{eq:PS_def}
\end{equation}
\noindent
where $k$ is the wavenumber and $\delta_{\rm D}$ is the Dirac delta function. We use discrete Fourier transforms of our estimator, $\tilde{\delta}_{\rm F}$, by following \citet{Khan_2024} (see their sec. 2 for more details). In particular, we estimate the $P_{21}$ in Eq.~\ref{eq:PS_def} with
\begin{equation}
    P_{21}\left(k_q\right)=\left(\frac{2\pi}{n\Delta\nu}\right)\langle\lvert\tilde{\delta}_{\rm F}\left(k_q\right)\rvert^2\rangle,\label{eq:PS_comp}
\end{equation}
\noindent
where $k_q=2\pi q/n\Delta\nu$ for $q=0,1,...,n-1$ and the total number of pixels in the 21-cm forest spectrum is $n$.

\begin{figure}
    \begin{minipage}{1\columnwidth}
 	  \centering
 	  \includegraphics[width=\linewidth]{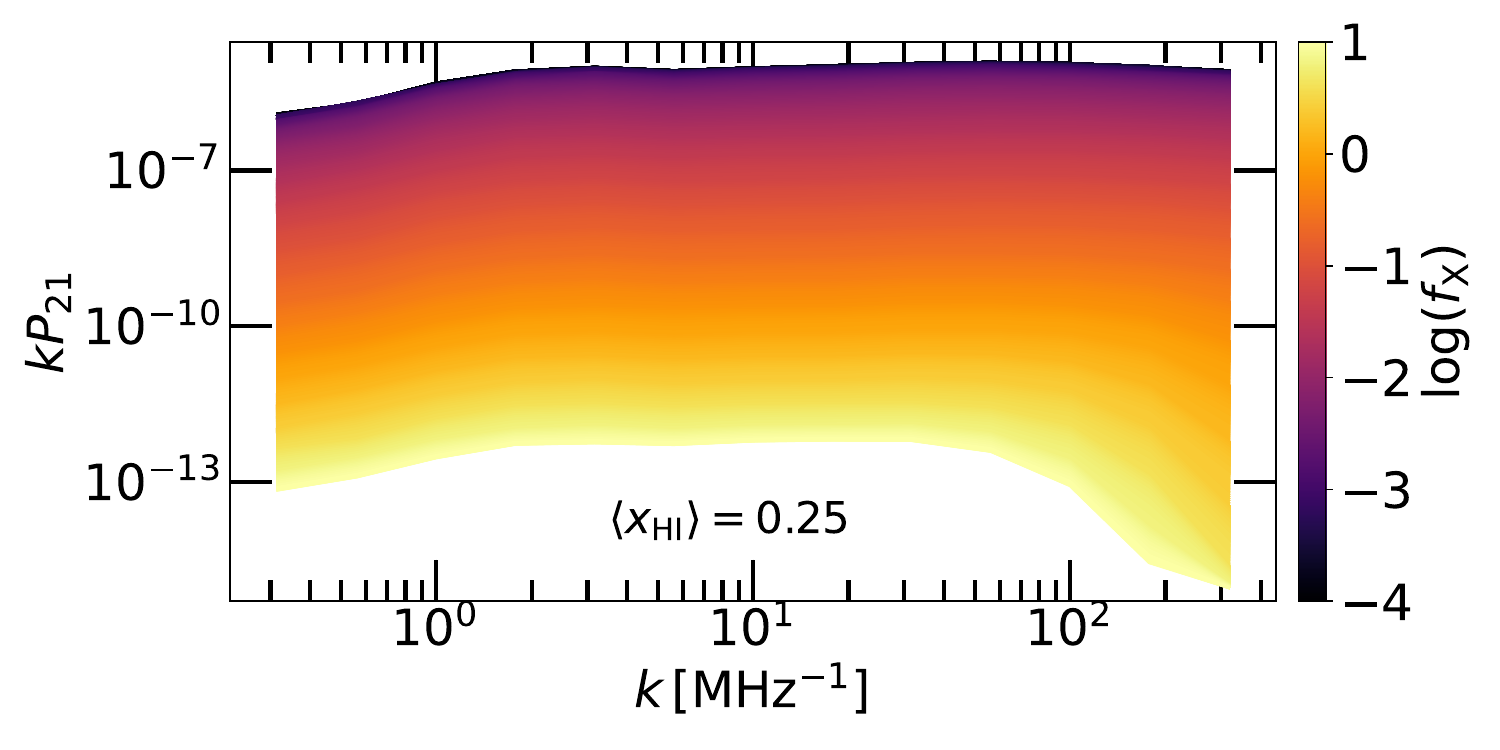}
	\end{minipage}
     \begin{minipage}{1\columnwidth}
 	  \centering
 	  \includegraphics[width=\linewidth]{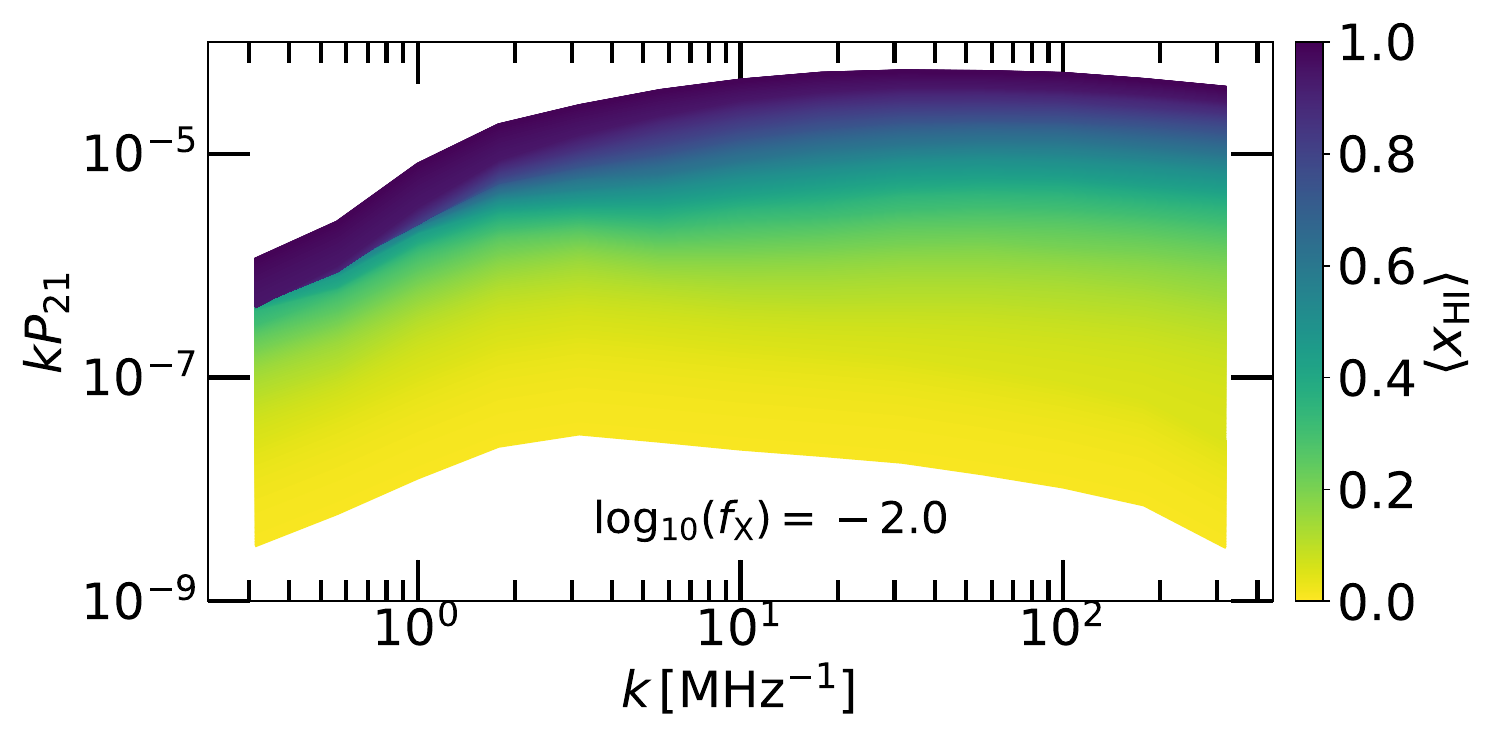}
	\end{minipage}
	\vspace{-0.3cm}
    \caption{\textit{Top panel}: Median 21-cm forest power spectrum, $P_{21}$ from 1000 realizations of LOS for fixed $\langle x_{\rm HI}\rangle=0.25$ and varying $\mathrm{log}_{10}f_{\rm X}$ (values shown in colorbar). \textit{Bottom panel}: Same as the top panel but with fixed $\mathrm{log}_{10}f_{\rm X}=-2$ and varying $\langle x_{\rm HI}\rangle$ .}
    \label{fig:1DPS_vsfXxHI_interpolate}
\end{figure}

We show an example of $kP_{21}$ in Fig.~\ref{fig:1DPS_noise} for our fiducial model (i.e. $\mathrm{log}_{10}f_{\rm X}=-2$ and $\langle x_{\rm HI}\rangle=0.25$). The solid orange curve and shaded region represent the median and the 68 per cent range of the $kP_{21}$ from 1000 LOS. Throughout the manuscript we have binned the $kP_{21}$ over the range of $10^{-0.5}\,\mathrm{MHz^{-1}}\leq k\leq 10^{2.5}\,\mathrm{MHz^{-1}}$ in the logarithmic $k$-bins of size \MNRASreply{$0.25$} of dex, unless stated otherwise. The shape of the $P_{21}$ broadly agrees with the literature \citep{Thyagaragan_2020,Shao_2023,Sun_2024}, particularly $kP_{21}$ increasing with $k$ at large scales and flattening at $k\gtrsim2\,\rm MHz^{-1}$. When comparing with previous works note that we present our results in terms of $kP_{21}$ instead of $P_{21}$. The dashed fuchsia curves mark the noise limit for the same observational setups as in Fig.~\ref{fig:noisyspectrum}. The noise limits increase almost linearly with  $k$ as expected from a Gaussian white noise. The slight deviation from the linear behaviour reflects the frequency dependent sensitivity of the telescope and flux normalization due to the intrinsic spectrum of the background source described in Sec.~\ref{sec:instrumental_features}. One can clearly see that the signal rises above these noise limits at large scales. Note that the noise limit for the SKA1-low is lower even if we assume 10 times shorter observational time.

\MNRASreplyreplyreply{Furthermore, we have tested the effect of the simulation resolution on our results. We find that the $kP_{21}$ is boosted by up to $9\%$ at $k\approx6\,\rm MHz^{-1}$ when the simulation resolution is increased to $512^3$ pixels. However, this difference is significantly less than the uncertainty arising from the sample variance. More details can be found in Appendix~\ref{app:convergence_test}.}

Besides the effect of instrumental features \MNRASreplyreplyreply{and simulation resolution} on the detectability of the 21-cm forest power spectrum, we also explore the effect of various physical mechanisms. The position of individual absorption features in the 21-cm forest in the redshift space can be shifted and their depth can be significantly boosted by the peculiar velocity of the gas \citep{Semelin_2016,Soltinsky_2021}. Both of these can potentially affect the $P_{21}$. We test this in Appendix~\ref{app:vpec} and find that while the $P_{21}$ is not affected at $k\lesssim20\,\rm MHz^{-1}$ (for our fiducial model of $\mathrm{log}_{10}f_{\rm X}=-2$, $\langle x_{\rm HI}\rangle=0.25$), at smaller scales the signal is boosted when we include $v_{\rm pec}$ in our calculation. However, the power spectrum that is affected by redshift space distortions significantly is either below the noise limit or even not accessible due to the limited spectral resolution.

\begin{figure*}
    \begin{minipage}{0.33\linewidth}
 	  \centering
 	  \includegraphics[width=\linewidth]{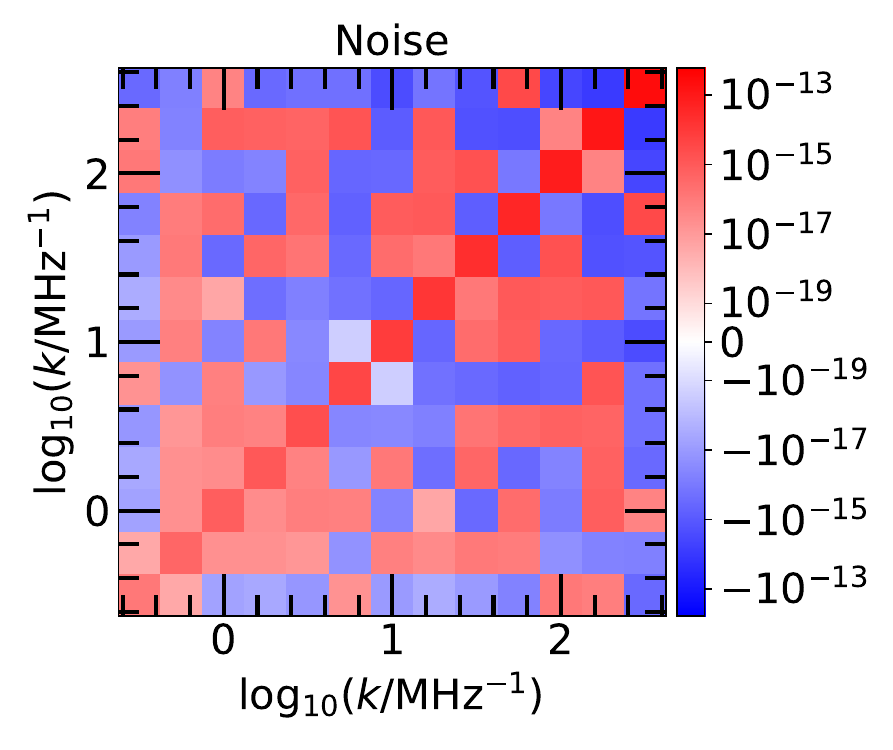}
	\end{minipage}
     \begin{minipage}{0.33\linewidth}
 	  \centering
 	  \includegraphics[width=\linewidth]{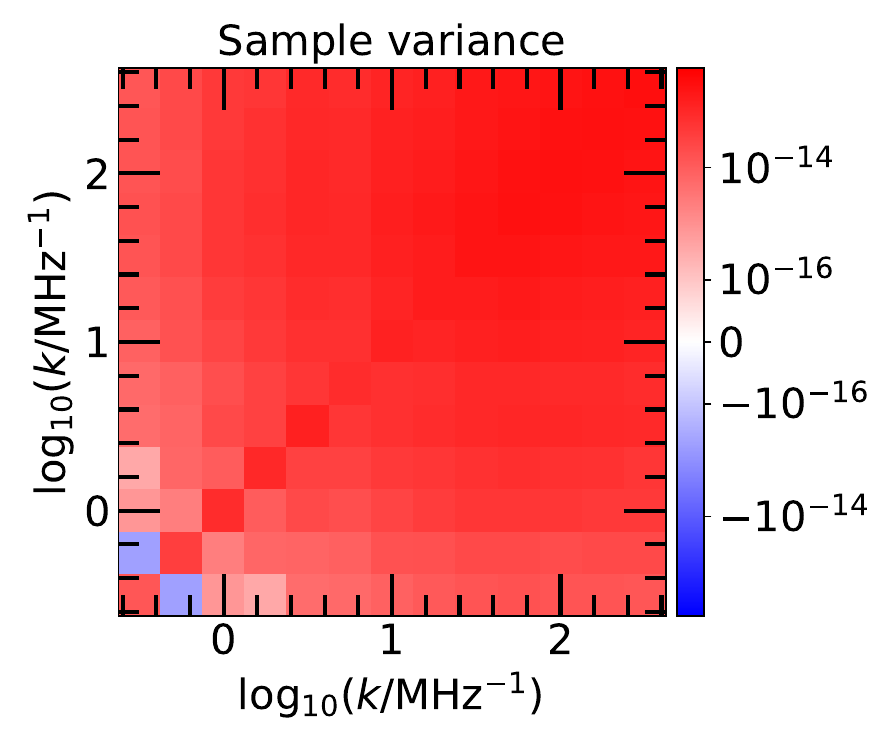}
	\end{minipage}
      \begin{minipage}{0.33\linewidth}
 	  \centering
 	  \includegraphics[width=\linewidth]{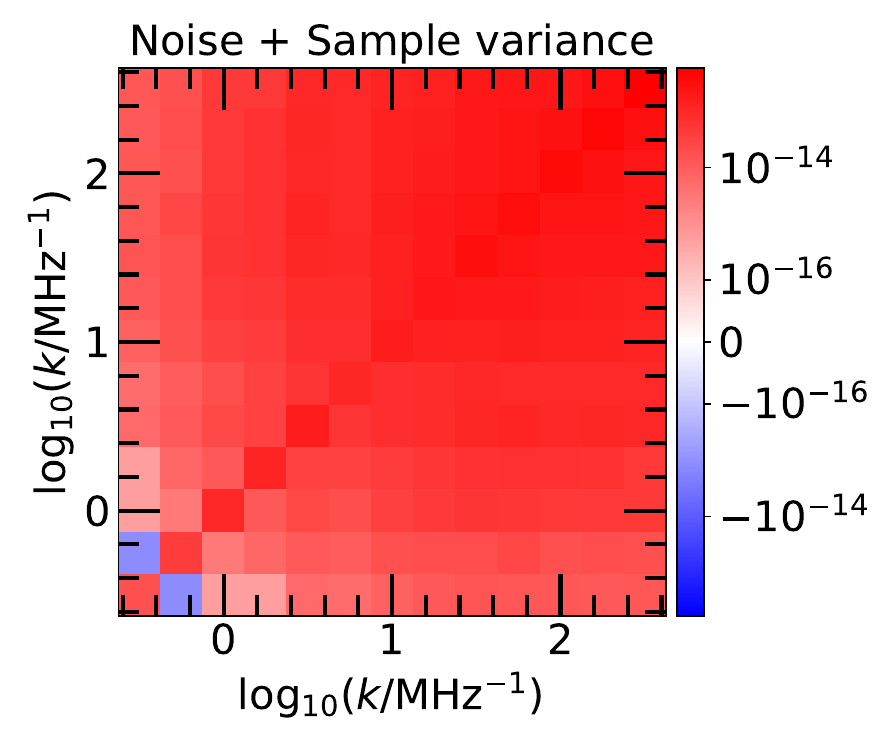}
	\end{minipage}
	\vspace{-0.3cm}
    \caption{Covariance matrix for the noise only (left panel), sample variance only (middle panel) and their combined effect on the mock observation (right panel) assuming an IGM with $\langle x_{\rm HI}\rangle=0.25$ and $\mathrm{log}_{10}f_{\rm X}=-2$ and $N_{\rm obs}=10$ observations of $200\,\rm cMpc$ spectra of a PSO J0309+27 like object (i.e. $S_{147}=64.2\, \rm mJy$ and $\alpha_{\rm R}=-0.44$) at $z=6$ with the uGMRT over $t_{\rm int}=500\, \rm hr$.}
    \label{fig:covariance_matrices}
\end{figure*}

To acquire the intuition of how the thermal and ionization state of the IGM affect the $P_{21}$, in Fig.~\ref{fig:1DPS_vsfXxHI_interpolate} we plot the median values of the $P_{21}$ from 1000 LOS with fixed value of one of the parameters governing the properties of the IGM while varying the other. In the top panel we fix $\langle x_{\rm HI}\rangle=0.25$ and the colormap shows the $P_{21}$ for different $\mathrm{log}_{10}f_{\rm X}$. As expected, the higher the $f_{\rm X}$, and hence the more preheating of the IGM by X-rays, the more suppressed the signal becomes. Similarly, the signal is suppressed if the IGM is more ionized as can be seen in the bottom panel of Fig.~\ref{fig:1DPS_vsfXxHI_interpolate}, in which we fix the $\mathrm{log}_{10}f_{\rm X}=-2$ and vary the $\langle x_{\rm HI}\rangle$ (colormap). The shape of the $P_{21}$ as a function of $k$ does not change significantly, however the slope changes at $k\gtrsim50\,\rm MHz^{-1}$ bin.


\section{Constraining the thermal and ionization state of the intergalactic medium}\label{sec:constraining_IGM}

Given this strong sensitivity of the $P_{21}$ on both the $\mathrm{log}_{10}f_{\rm X}$ and the $\langle x_{\rm HI}\rangle$, here we explore the potential of the $P_{21}$ measurements to constrain both the thermal and ionization state of the IGM at $z=6$. \MNRASreply{It is important to note that the 21-cm line is one of the few probes of the temperature of cold and neutral IGM regions.} Similarly to \citet{Shao_2023}, who took the advantage of how differently the amplitude and shape of the 21-cm forest power spectrum\footnote{Note that their power spectrum was computed from the 21-cm forest spectrum which was given in terms of differential brightness temperature as opposed to the transmitted flux as was done in this study.} vary with the efficiency of X-ray background radiation and the mass of warm dark matter particles to constrain these two quantities, we will explore the potential of the $P_{21}$ to constrain the $\mathrm{log}_{10}f_{\rm X}$ and the $\langle x_{\rm HI}\rangle$. For this we will use Bayesian statistical methods described in Sec.~\ref{sec:MCMC}. In Sec.~\ref{sec:results} we discuss implications of our results.

\subsection{Statistical inference procedure}\label{sec:MCMC}

In what follows we assume a measurement of 10 radio spectra of $22.1\,\rm MHz$ bandwidth each which corresponds to $200\,\rm cMpc$ long LOS at $z=6$. As mentioned before, the uGMRT band-2 has a total bandwidth of $150\,\rm MHz$, and therefore observation of longer spectra than considered here are possible. However, we are limited by the type of the simulation and its box size. We assume identical intrinsic spectra of the background sources for all 10 LOS, particularly $S_{147}=64.2\,\rm mJy$ and $\alpha_{\rm R}=-0.44$. This corresponds to the PSO J0309+27 which is one of the brightest RLQSOs at $z>5.5$ currently known as one can see in the bottom panel of Fig.~\ref{fig:RLQdist}. While this might be an optimistic observational requirement, we remind the reader that \citet{Niu_2024} predicts more than 50 RLQSOs of $S_{147}=100\,\rm mJy$ at $z>5.5$ in the whole sky. For this \MNRASreply{kind of} observation we will define $\langle P_{21}\rangle_{10}$ which we construct by randomly selecting 10 synthetic 21-cm forest spectra, calculating their $P_{21}$ and taking the mean at each $k$ bin.

As explained above, our model is described by the X-ray background radiation efficiency and mean neutral hydrogen fraction. Hence, let $\bm{\theta}=\{\mathrm{log}_{10}f_{\rm X},\langle x_{\rm HI}\rangle\}$. We assume uniform prior distributions, $\mathcal{P}(\bm{\theta})$, of the parameters. For the X-ray background radiation efficiency we have chosen a range of $\mathrm{log}_{10}f_{\rm X}=[-4,1]$. For the neutral hydrogen fraction we consider the whole physical range of $\langle x_{\rm HI}\rangle=[0,1]$ at $\mathrm{log}_{10}f_{\rm X}<-0.6$ and $\langle x_{\rm HI}\rangle=[0,1-0.1125(\mathrm{log}_{10}f_{\rm X}+0.6)]$ at $\mathrm{log}_{10}f_{\rm X}\geq-0.6$. The latter range is selected this way because in our simulations the IGM is at least partially ionized by the X-rays even if we set the background ionization efficiency to 0 as explained in Sec.~\ref{sec:IGM_sims}. The highest $\langle x_{\rm HI}\rangle$ in these models has slightly higher value than the one given by the relation for the upper bound on $\langle x_{\rm HI}\rangle$.

To constrain these parameters, we follow Bayesian procedure with Gaussian likelihood defined as
\begin{equation}
    \mathcal{L}(P_{21}\lvert\bm{\theta})=\frac{1}{\sqrt{\mathrm{det\textbf{C}}}}\exp\left(-\frac{1}{2}\mathrm{\textbf{d}^T\textbf{C}^{-1}\textbf{d}}\right),\label{eq:likelihood_covar}
\end{equation}
\noindent
where $\textbf{C}$ is the covariance matrix which is constructed from an ensemble of $10^4$ $\langle P_{21}\rangle_{10}$. Firstly, we consider the effect of the telescope noise by computing $\textbf{C}$ from the noisy radio spectra without the 21-cm forest signal. Such covariance matrix, for an observation by the uGMRT over $t_{\rm int}=500\,\rm hr$ \MNRASreply{per source} pointing at \MNRASreply{quasars} with $S_{147}=64.2\,\rm mJy$ and $\alpha_{\rm R}=-0.44$ (this corresponds to the upper dashed fuchsia line in Fig.~\ref{fig:1DPS_noise}), is shown in the left panel of Fig.~\ref{fig:covariance_matrices}. This shows a relatively weak (anti-)correlation between various $k$ bins. In the middle panel of the same figure we show $\textbf{C}$ calculated from an ensemble \MNRASreply{of $\langle P_{21}\rangle_{10}$ incorporating only the 21-cm forest signal (not incorporating the telescope noise), which we define as $\langle P_{21}^{\rm sim}\rangle_{10}(\bm{\theta})$, i.e. simulated/model $\langle P_{21}\rangle_{10}$.} This is based on our fiducial IGM model with $\bm{\theta}=[-2,0.25]$. This panel shows the effect of the sample variance which is orders of magnitude stronger than the effect of the telescope noise. 

\begin{figure*}
    \begin{minipage}{0.66\linewidth}
 	  \centering
 	  \includegraphics[width=\linewidth]{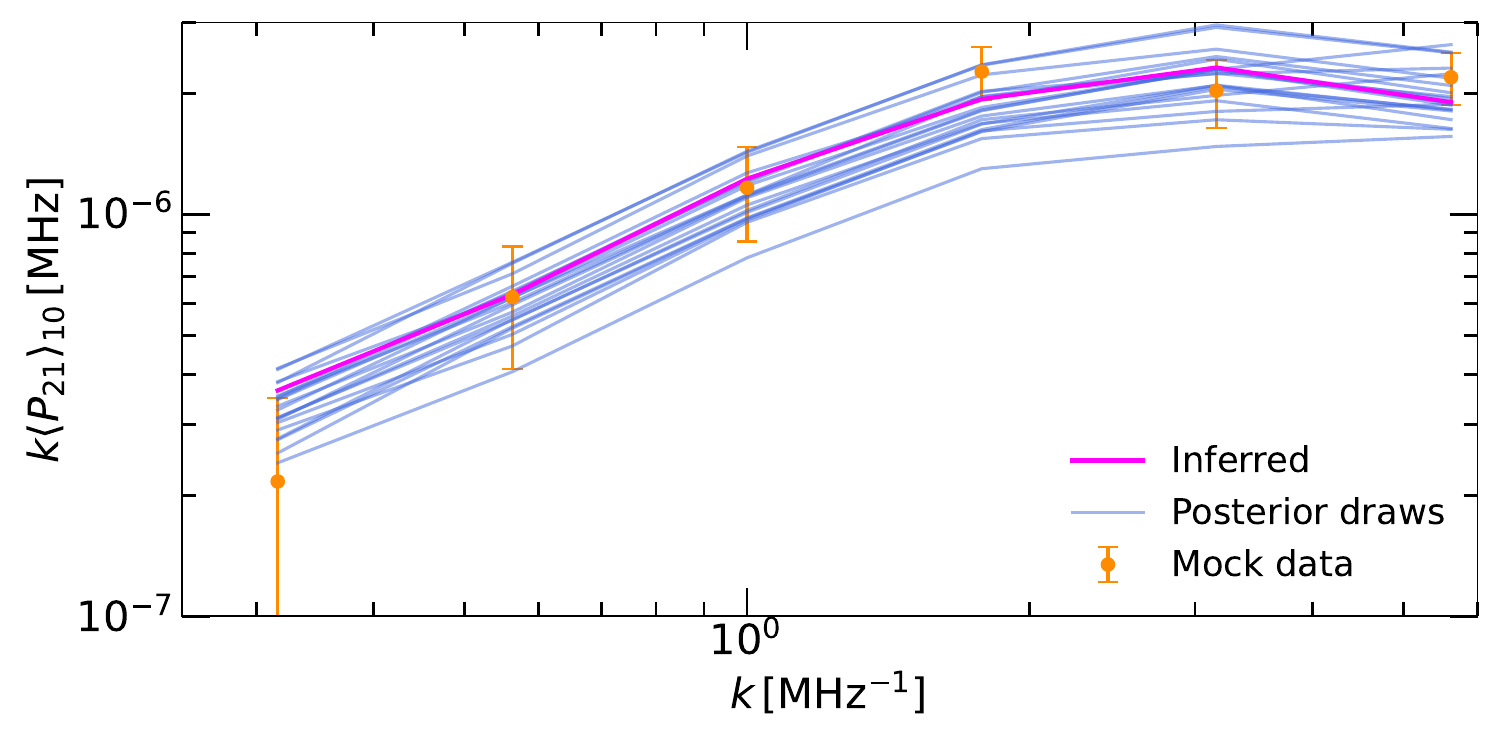}
    \end{minipage}
    \begin{minipage}{0.33\linewidth}
 	  \centering
 	  \includegraphics[width=\linewidth]{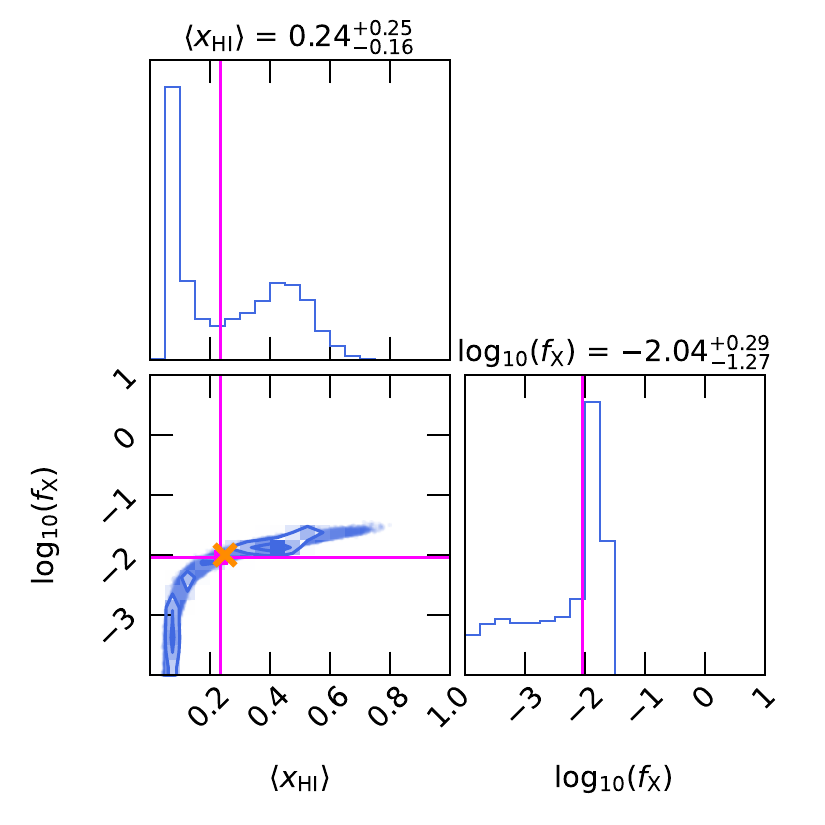}
    \end{minipage}
    \vspace{-0.3cm}
    \caption{Bayesian inference of $\mathrm{log}_{10}f_{\rm X}$ and $\langle x_{\rm HI}\rangle$ from a mock observation of $S_{147}=64.2\,\rm mJy$ and $\alpha_{\rm R}=-0.44$ quasar by uGMRT for $500\,\rm hr$. The true values of the IGM parameters are $\mathrm{log}_{10}f_{\rm X}=-2$, $\langle x_{\rm HI}\rangle=0.25$. \textit{Left panel}: The $\langle P_{21}\rangle_{10}$ calculated from a spectrum including signal and noise. The orange data points indicate a mock observation of the 1D 21-cm forest power spectrum, $k\langle P_{21}\rangle_{10}$. The blue curves correspond to the $k\langle P_{21}\rangle_{10}$ randomly drawn from posterior acquired from the MCMC sampling. The fuchsia curve shows the $k\langle P_{21}\rangle_{10}$ for the inferred values of $\mathrm{log}_{10}f_{\rm X}=-2.04$, $\langle x_{\rm HI}\rangle=0.23$. \textit{Right panels}: Corner plot of the posterior distributions computed from the MCMC sampling (blue shaded region and blue curves). The true model parameters are indicated by the orange cross. Our analyses jointly recovers both parameters values of $\mathrm{log}_{10}f_{\rm X}=-2.03^{+0.25}_{-1.19}$ and $\langle x_{\rm HI}\rangle=0.24^{+0.20}_{-0.17}$ as indicated by the fuchsia square and lines.}
    \label{fig:MCMC_uGMRT_500h}
\end{figure*}

To construct a mock observation, $\langle P_{21}^{\rm mock}\rangle_{10}$, we need to incorporate both the effect of the telescope noise and sample variance. We do so by using \MNRASreply{10 randomly} selected synthetic 21-cm forest spectra including instrumental features (e.g., black curves in Fig.~\ref{fig:noisyspectrum}). An example of a single $\langle P_{21}^{\rm mock}\rangle_{10}$ is presented as orange points in Fig.~\ref{fig:MCMC_uGMRT_500h} considering the same observational setup and the same IGM model as above. The errorbars indicate $1\sigma$ uncertainty computed from an ensemble of $10^4$ $\langle P_{21}^{\rm mock}\rangle_{10}$ (hence, incorporating uncertainty from both the telescope noise and sample variance). We compute the $\textbf{C}$ shown in the right panel of Fig.~\ref{fig:covariance_matrices} from this ensemble of \MNRASreply{$10^4$} $\langle P_{21}^{\rm mock}\rangle_{10}$. This is very similar to the middle panel, and hence we \MNRASreply{infer} that the IGM parameter inference from the $P_{21}$ at $z\approx6$ will be dominated by the sample variance rather than the telescope noise. 

The matrix $\textbf{d}=\langle P_{21}^{\rm mock}\rangle_{10}-\langle P_{21}^{\rm sim}\rangle_{10}(\bm{\theta})$. With all of the above in place, we can utilize the Bayes theorem
\begin{equation}
    P(\bm{\theta}\lvert P_{21}^{\rm mock})\propto\mathcal{L}(P_{21}^{\rm mock}\lvert\bm{\theta})\mathcal{P}(\bm{\theta}),\label{eq:bayes_theorem}
\end{equation}
\noindent
where $P(\bm{\theta}\lvert P_{21}^{\rm mock})$ is the posterior probability distribution of the parameter values $\bm{\theta}$ given the mock observation $\langle P_{21}^{\rm mock}\rangle_{10}$. We implement the MCMC \citep{Goodman_2010} using the \textsc{emcee} software \citep{Foreman-Mackey_2013_emcee} to sample the \MNRASreply{$\langle P_{21}^{\rm sim}\rangle_{10}(\bm{\theta})$}\footnote{We run 64 Markov chains over 100000 steps for each parameter. First 500 steps are disregarded given that the autocorrelation time is $<350$. The runs are initiated at the parameters initial guesses given by maximising the $\mathcal{L}(P_{21}^{\rm mock}\lvert\bm{\theta})$.}. Therefore, we can neglect the normalization. \MNRASreply{Note that the $\langle P_{21}^{\rm sim}\rangle_{10}(\bm{\theta})$ is calculated at any $\bm{\theta}$ using linear interpolation depending on the MCMC sampling as long as the $\bm{\theta}$ is within the ranges defining the $\mathcal{P}(\bm{\theta})$.}

\subsection{Results}\label{sec:results}

Here we implement the procedure described in Sec.~\ref{sec:MCMC} to explore the potential of the $P_{21}$ to constrain both the thermal and ionization state of the IGM. However, before we discuss the results of this procedure, we note that while the sample variance is the dominating effect in the parameter inference, the telescope noise also contributes to increasing the uncertainty on the inferred values of the IGM properties. Therefore, in our analyses we neglect the \MNRASreply{$k>8.5\,\rm MHz^{-1}$} bins at which, as one can see in Fig.~\ref{fig:1DPS_noise}, the fiducial IGM model $P_{21}$ is below or close to the noise limit. This significantly improves the parameter inference. After inspecting the $P_{21}$ behaviour at low $k$, \MNRASreply{we suspect that adding bins with $k<0.32\,\rm MHz^{-1}$} might improve our analyses. However, longer 21-cm forest spectra are required to do so.

The results of our Bayesian inference procedure are presented in Fig.~\ref{fig:MCMC_uGMRT_500h} where we used $\langle P_{21}^{\rm mock}\rangle_{10}$ assuming $\bm{\theta}=[-2,0.25]$ and an $t_{\rm int}=500\,\rm hr$ \MNRASreply{per source} observation by uGMRT. In the left panel, the blue curves represent 20 randomly drawn $\langle P_{21}^{\rm sim}\rangle_{10}$ from the posterior distribution from the MCMC. They fit the errorbars of the mock observation (orange points) reasonably well. Our analyses infers the $\langle P_{21}^{\rm sim}\rangle_{10}$ shown as the fuchsia curve. This corresponds to the inferred $\mathrm{log}_{10}f_{\rm X}=-2.03_{-1.19}^{+0.25}$ and $\langle x_{\rm HI}\rangle=0.24_{-0.17}^{+0.20}$ indicated with the fuchsia square and horizontal and vertical lines in the corner plot on the right in Fig.~\ref{fig:MCMC_uGMRT_500h}. The true values of the IGM parameters (orange cross in the corner plot) lie very close to the mean inferred values. However, only upper limit on $\mathrm{log}_{10}f_{\rm X}$ and lower limit on $\langle x_{\rm HI}\rangle$ is acquired, as shown in the marginalized posterior distributions. Note that the $P_{21}$ does not change significantly when one decreases $\mathrm{log}_{10}f_{\rm X}$ below $\approx-3$ which is shown in the top panel of Fig.~\ref{fig:1DPS_vsfXxHI_interpolate}. This range is smaller than the errorbars on the mock $\langle P_{21}^{\rm mock}\rangle_{10}$ and as a consequence it is difficult for our analyses to distinguish between different $\mathrm{log}_{10}f_{\rm X}$ in this parameter space. This results in a vertical tail of the posterior distribution in Fig.~\ref{fig:MCMC_uGMRT_500h}.

\begin{figure}
    \begin{minipage}{1\columnwidth}
 	  \centering
 	  \includegraphics[width=\linewidth]{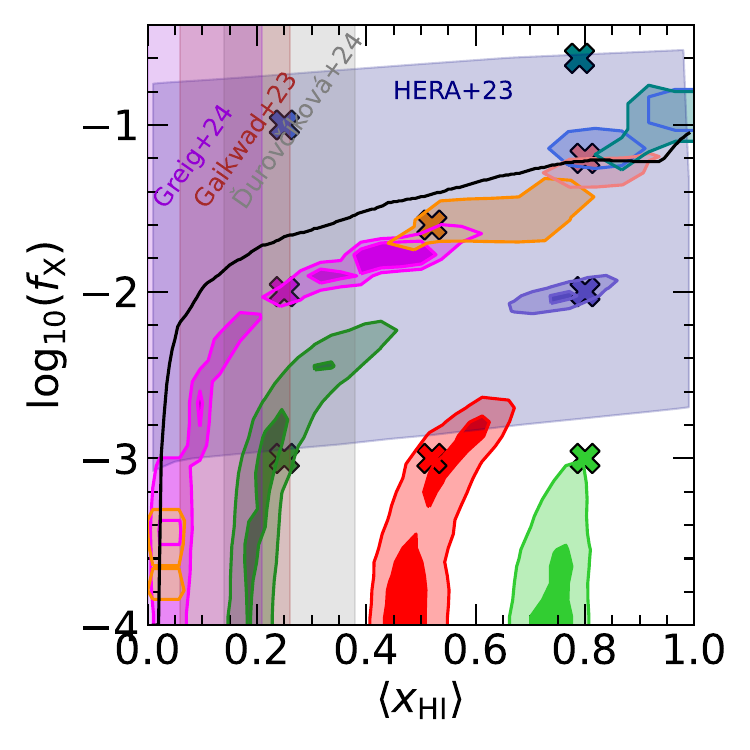}
	\end{minipage}
    \vspace{-0.3cm}
    \caption{Posterior distributions corresponding to 1 and 2$\sigma$ confidence regions indicated by contours for different true values (different colours) of the $\mathrm{log}_{10}f_{\rm X}$ and $\langle x_{\rm HI}\rangle$ indicated by the crosses. This shows the constraining power of the 21-cm forest power spectrum for an observation of $S_{147}=64.2\,\rm mJy$ and $\alpha_{\rm R}=-0.44$ quasar by uGMRT for $500\,\rm hr$. For comparison, \Lya~forest measurements of the $\langle x_{\rm HI}\rangle$ at $z\approx6$, particularly $\langle x_{\rm HI}\rangle=0.17^{+0.09}_{-0.11}$ \citep[brown shaded region,][]{Gaikwad_2023}, $\langle x_{\rm HI}\rangle=0.21^{+0.17}_{-0.07}$ \citep[grey shaded region,][]{Durovcikova_2024} and $\langle x_{\rm HI}\rangle<0.21$ \citep[purple shaded region,][]{Greig_2024} \MNRASreply{is shown}. The dark blue shaded region encompasses the $\mathrm{log}_{10}f_{\rm X}$ which in our modelling corresponds to the spin temperature range inferred from the 21-cm power spectrum limits from \citet{Hera_2023}. \MNRASreply{The solid  black curve indicates the disfavoured parameter space (lower $\mathrm{log}_{10}f_{\rm X}$, higher $\langle x_{\rm HI}\rangle$) in the case of a null-detection. Note that in the cases outside of this region the telescope noise dominates over the signal and as a consequence the posterior distribution deviates away from the true value.}}
    \label{fig:multi_param_estimation_uGMRT_500h}
\end{figure}

The constraining power of the $P_{21}$ mock measurement with the same observational setup as shown in Fig.~\ref{fig:MCMC_uGMRT_500h} over the whole considered parameter space is summarized in Fig.~\ref{fig:multi_param_estimation_uGMRT_500h}. We show the $1$ and $2\sigma$ credible intervals as the dark and light shaded regions with the crosses of the same colour indicating the true values used to generate mock observations. The more the IGM is neutral the tighter constraints can be acquired, as expected. Moreover, while colder IGM (i.e. lower $\mathrm{log}_{10}f_{\rm X}$) would lead to tighter constraints, if the $\mathrm{log}_{10}f_{\rm X}$ becomes too low, the credible intervals become elongated towards even lower values due to the diminutive changes in the $P_{21}$ described above. However, if the $\mathrm{log}_{10}f_{\rm X}$ is low, one can infer tighter constraints on the ionization state of the IGM. 

On the other hand, if the IGM has experienced a significant heating by the X-ray background radiation and/or is significantly ionized such as in the case of the blue and teal crosses ($\mathrm{log}_{10}f_{\rm X}=-1$ and $-0.6$, and $\langle x_{\rm HI}\rangle=0.25$ and $0.79$, respectively), \MNRASreply{the telescope noise dominates over the signal and the $\langle P_{21}^{\rm mock}\rangle_{10}$ is boosted significantly. Given that the $\langle P_{21}^{\rm sim}\rangle_{10}(\bm{\theta})$ does not contain this boost from the telescope noise, the MCMC procedure is forced to find the best fitting $\langle P_{21}^{\rm sim}\rangle_{10}(\bm{\theta})$ from stronger signal, and hence from lower $\mathrm{log}_{10}f_{\rm X}$ and/or higher $\langle x_{\rm HI}\rangle$. This is reflected in some of the posterior distributions deviating from the true parameter values of $\bm{\theta}$. Prime examples of this are the mock observations from the model of $\mathrm{log}_{10}f_{\rm X}=-1$, $\langle x_{\rm HI}\rangle=0.25$ (blue cross and shaded region) and $\mathrm{log}_{10}f_{\rm X}=-0.6$, $\langle x_{\rm HI}\rangle=0.79$ (teal cross and shaded region).}


\MNRASreply{Naturally, in the case of telescope noise dominating over the signal one would consider this to be a null-detection of the signal. We define a null-detection of the $\langle P_{21}\rangle_{10}$ as the case in which the $\langle P_{21}^{\rm sim}\rangle_{10}$ in at least one of the $k$-bins at $<8.5\,\rm MHz^{-1}$ is below the 1D power spectrum of the noise only. However, similarly to \citet{Soltinsky_2021} we suggest that a null-detection could be translated into joint lower limit on the $\mathrm{log}_{10}f_{\rm X}$ and upper limit on the $\langle x_{\rm HI}\rangle$. We show the $1\sigma$ limit on these parameters as the solid black curve in Fig.~\ref{fig:multi_param_estimation_uGMRT_500h}. To acquire this curve we generate an ensemble of $10^4$ $\langle P_{21}^{\rm sim}\rangle_{10}(\bm{\theta})$ and find $\bm{\theta}$ at which $32\%$ of them result in a null-detection according to the definition described above. In other words, if the observation of 10 RLQSO by the uGMRT over $500\,\rm hr$ each results in a null-detection, one can disfavour the parameter space below this solid black curve.} When comparing with the measurements of $\langle x_{\rm HI}\rangle$ from the \Lya~forest observations, particularly \citet{Gaikwad_2023} (brown shaded \MNRASreply{vertical} region), \citet{Greig_2024} (purple shaded \MNRASreply{vertical} region) and \citet{Durovcikova_2024} (grey shaded \MNRASreply{vertical} region), one can see that the $\langle P_{21}\rangle_{10}$ has a potential to constrain the $\langle x_{\rm HI}\rangle$ even more than currently available measurements. This is even more pronounced in the case of $f_{\rm X}$. The $\mathrm{log}_{10}f_{\rm X}$ corresponding to $15.6\,\mathrm{K}\leq T_{\rm K}\leq656.7\,\mathrm{K}$, which is the spin temperature range inferred from the 21-cm power spectrum measurements by \citet{Hera_2023}, in our modelling is shown as a dark blue shaded region. It is obvious that according to the current observations the $f_{\rm X}$ is allowed to range over few orders of magnitude which could be significantly constrained with the $\langle P_{21}\rangle_{10}$ observations.

\begin{figure*}
    \begin{minipage}{0.66\linewidth}
 	  \centering
 	  \includegraphics[width=\linewidth]{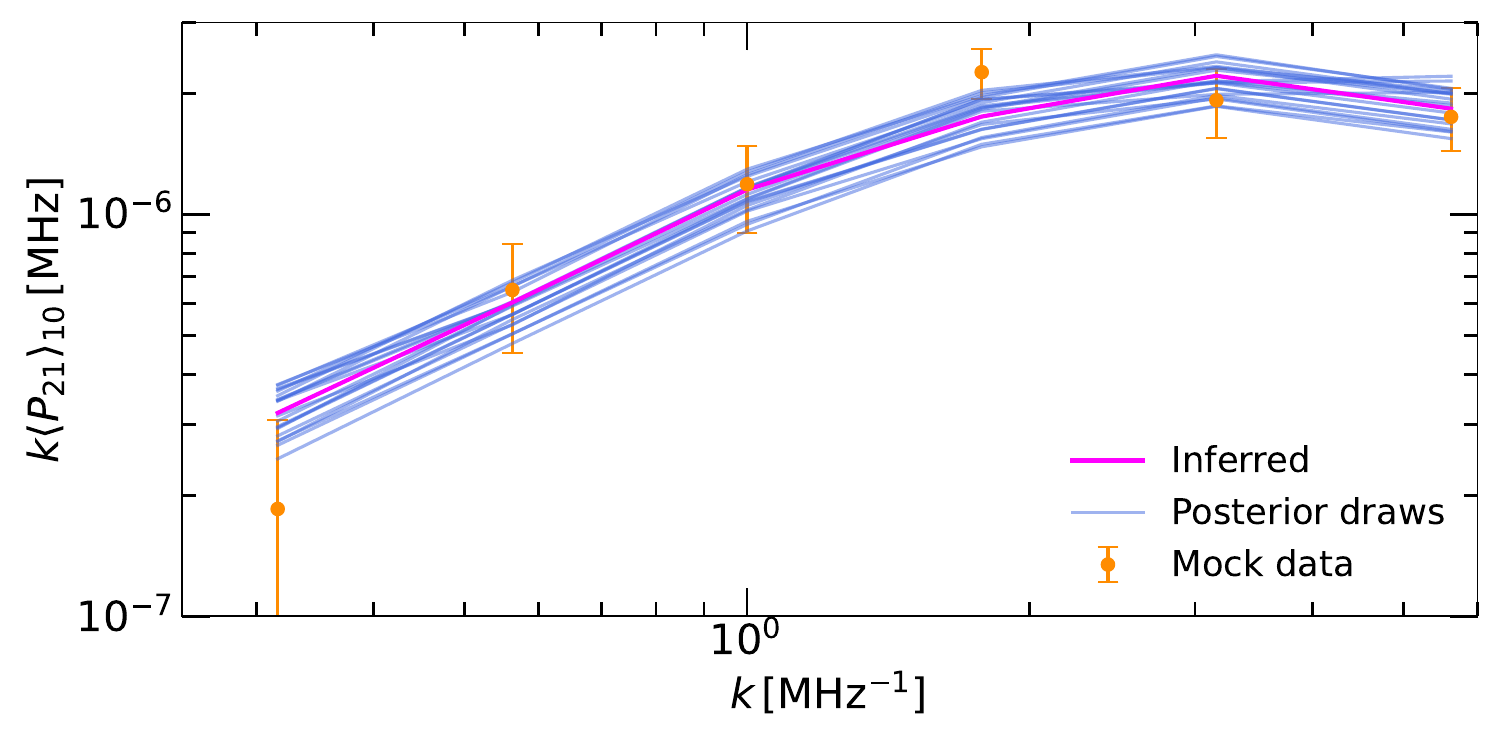}
    \end{minipage}
    \begin{minipage}{0.33\linewidth}
 	  \centering
 	  \includegraphics[width=\linewidth]{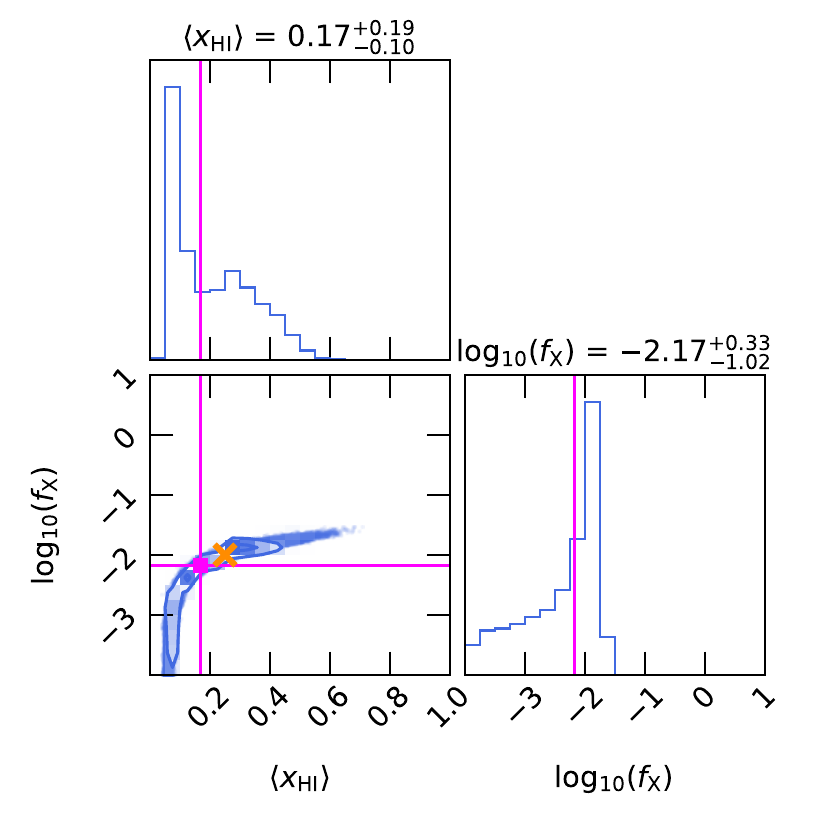}
    \end{minipage}
    \vspace{-0.3cm}
    \caption{Same as Fig.~\ref{fig:MCMC_uGMRT_500h} but for a mock observation by SKA1-low ant $t_{\rm int}=50\,\rm hr$. The jointly recovered parameters values are $\mathrm{log}_{10}f_{\rm X}=-2.22^{+0.36}_{-1.03}$ and $\langle x_{\rm HI}\rangle=0.15^{+0.18}_{-0.08}$.}
    \label{fig:MCMC_SKA1-low}
\end{figure*}

We repeat the same analyses for a mock observation by the SKA1-low with $t_{\rm int}=50\,\rm hr$ \MNRASreply{per source} and observing \MNRASreply{quasars with the same properties} at $z=6$ ($S_{\rm min}=64.2\,\rm mJy$, $\alpha_{\rm R}=-0.44$). We show the results in Fig.~\ref{fig:MCMC_SKA1-low} where we again assumed a mock observation of the IGM with $\mathrm{log}_{10}f_{\rm X}=-2$ and $\langle x_{\rm HI}\rangle=0.25$ (orange points in the left panel). One can see in the corner plot on the right that the posterior distribution does not differ much from the case of the observation by the uGMRT and $t_{\rm int}=500\,\rm hr$ \MNRASreply{per source}. This is most likely caused by the sample variance effect dominating over the telescope noise as explained in Section~\ref{sec:MCMC}. Moreover, the inferred values, $\mathrm{log}_{10}f_{\rm X}=-2.22^{+0.36}_{-1.03}$ and $\langle x_{\rm HI}\rangle=0.15^{+0.18}_{-0.08}$, indicated by fuchsia deviate significantly more from the true values (orange cross) than in the previous case. This is caused by the $k=0.562\,\rm MHz^{-1}$ bin which is difficult to reproduce with the $\langle P_{21}^{\rm sim}\rangle_{10}$ while fitting the other $k$ bins. However, excluding the lowest $k$ bin does not improve the constraining power of the $\langle P_{21}\rangle_{10}$ on $\bm{\theta}$. On the other hand, the constraints on the IGM parameters are slightly tighter. This is also shown in Fig.~\ref{fig:multi_param_estimation_SKA1-low} in which the posterior distributions can encompass the true values at higher $\mathrm{log}_{10}f_{\rm X}$ relatively to a case of the uGMRT observational setup considered previously. \MNRASreply{In addition,} if the SKA1-low measurements of the $\langle P_{21}\rangle_{10}$ will result in a null-detection, it will lead to a tighter lower limit on the thermal state of the IGM at $z=6$ even with 10 times shorter observational time. \MNRASreply{For instance, at $\langle x_{\rm HI}\rangle=0.25$ and $0.50$ the lower limit on the X-ray background radiation efficiency would increase to $\mathrm{log}_{10}f_{\rm X}\gtrsim-1.42$ and $\gtrsim-1.20$ for the SKA1-low, $50\,\rm hr$ per source observation, respectively, as compared to the $\mathrm{log}_{10}f_{\rm X}\gtrsim-1.67$ and $\gtrsim-1.43$ in the case of the observation by the uGMRT, $500\,\rm hr$ per source.}

Finally, we consider an observation by the uGMRT over $t_{\rm int}=50\,\rm hr$ \MNRASreply{per source} with the aim to test what is possible with currently operational instruments and is relatively not expensive. In this case, the telescope noise is 10 times larger (in terms of $kP_{21})$ than in the case of $500\,\rm hr$ \MNRASreply{per source} with the uGMRT. Consequently, the MCMC results in posterior distributions that include the true values (crosses) within $2\sigma$ only in the cases of very cold and significantly neutral IGM, as can be seen in Fig.~\ref{fig:multi_param_estimation_uGMRT_50h}. \MNRASreply{Furthermore}, the disfavoured parameter space with a null-detection utilizing such observations is much smaller. \MNRASreply{Despite this, the limits on the ionization state of the gas for a very cold IGM can reach $\langle x_{\rm HI}\rangle\lesssim0.18$ which is tighter than the currently available constraints on the $\langle x_{\rm HI}\rangle$ from the \Lya~forest, particularly from \citet{Gaikwad_2023}, \citet{Greig_2024} and \citet{Durovcikova_2024}. The limits of $\mathrm{log}_{10}f_{\rm X}\gtrsim-2.68$ and $\gtrsim-2.07$ if $\langle x_{\rm HI}\rangle=0.25$ and $0.50$, respectively. Also, note that we show only $1\sigma$ and $2\sigma$ credible intervals in Fig.~\ref{fig:multi_param_estimation_uGMRT_500h},~\ref{fig:multi_param_estimation_SKA1-low} and~\ref{fig:multi_param_estimation_uGMRT_50h}. We have checked that the full posterior distributions encapsulate their corresponding true values as long as they are within the solid black curve (even for the case of $\mathrm{log}_{10}f_{\rm X}=-3.2$, $\langle x_{\rm HI}\rangle=0.25$ (dark green cross) and $\mathrm{log}_{10}f_{\rm X}=-2$, $\langle x_{\rm HI}\rangle=0.8$ (purple cross)).}


\section{Conclusions}\label{sec:conclusions}


\begin{figure}
    \begin{minipage}{1\columnwidth}
 	  \centering
 	  \includegraphics[width=\linewidth]{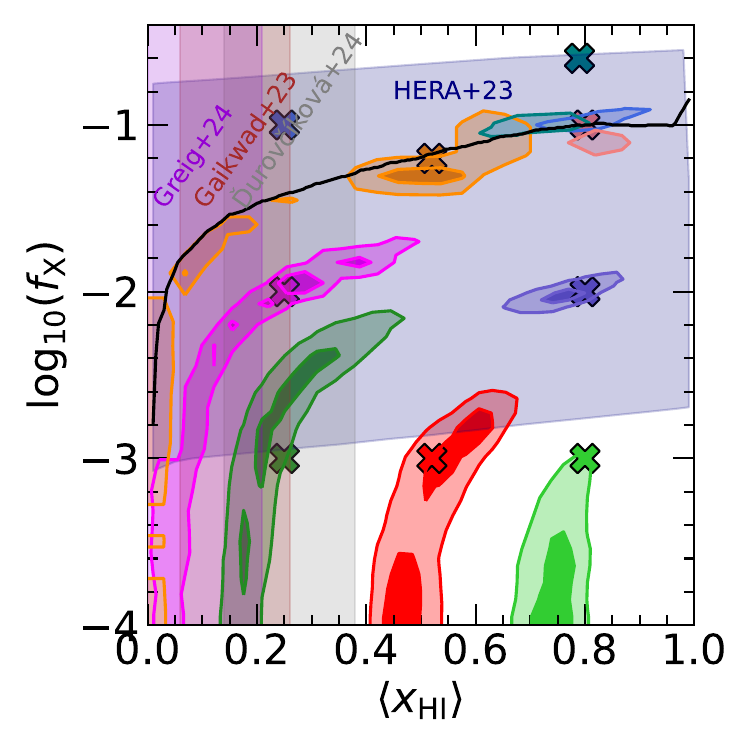}
	\end{minipage}
    \vspace{-0.3cm}
    \caption{Same as Fig.~\ref{fig:multi_param_estimation_uGMRT_500h} but for a mock observation by SKA1-low over $50\,\rm hr$.}
    \label{fig:multi_param_estimation_SKA1-low}
\end{figure}

\begin{figure}
    \begin{minipage}{1\columnwidth}
 	  \centering
 	  \includegraphics[width=\linewidth]{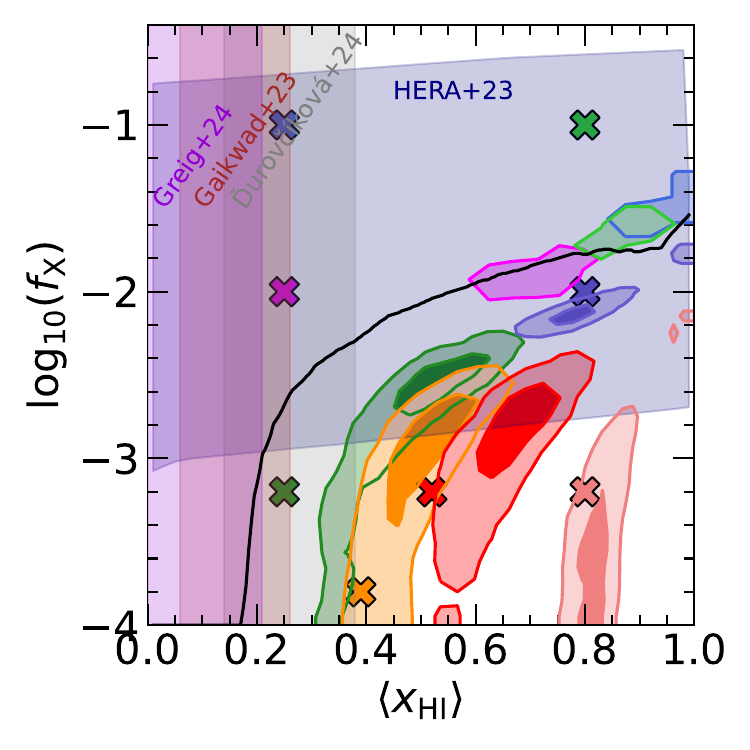}
	\end{minipage}
    \vspace{-0.3cm}
    \caption{Same as Fig.~\ref{fig:multi_param_estimation_uGMRT_500h} and~\ref{fig:multi_param_estimation_SKA1-low} but for a mock observation by uGMRT over $50\,\rm hr$.}
    \label{fig:multi_param_estimation_uGMRT_50h}
\end{figure}

In this study we have explored the detectability of the 21-cm forest signal with focus on currently available options, namely targeting already confirmed radio-loud quasars at $z\sim6$ with the uGMRT, but we also made forecasts for the SKA1-low telescope. We forward-modelled the 21-cm forest spectra using semi-numerical simulations of the IGM and incorporated several instrumental features. By considering a direct and statistical detection of the 21-cm forest signal and utilizing Bayesian techniques to constrain the thermal and ionization state of the IGM we found the following:

\begin{itemize}
    \item In the case of late-end reionization and only modest IGM preheating by the X-ray background radiation, particularly $\langle x_{\rm HI}\rangle=0.25$ and $\mathrm{log}_{10}f_{\rm X}=-2$, the SNR achieved by the uGMRT (SKA1-low) observing for $500\,\rm hr$ ($50\,\rm hr$) is $0.37$ ($1.2$). Therefore, rather than aiming for a direct detection of individual absorption features we suggest that a statistical detection of the 21-cm forest is more plausible.
    \item One of the statistics we considered in this study is the differential number density of the 21-cm forest flux at each frequency channel (or pixel). We propose that a discrepancy between the distribution of the $F_{21}$ of observed spectra and the noise distribution (normal distribution in our case), particularly at $F_{21}<1$, can be considered as a statistical detection of the signal. Considering the same IGM model as in the previous point and using two-sided Kolmogorov-Smirnov test, we have found that while such detection was achieved only in $\approx15\%$ of our synthetic spectra when observed by the uGMRT over $50\,\rm hr$, the fraction of detections increases greatly when either increasing the observational time by a factor of 10 or by using SKA1-low instead.
    \item Another promising strategy to acquire a detection of the 21-cm forest is the 1D power spectrum calculated from the normalized 21-cm forest flux, $P_{21}$. As can be seen in Fig.~\ref{fig:1DPS_noise}, the signal is above the noise limit at scales of \MNRASreply{$k\lesssim8.5\,\rm MHz^{-1}$ ($k\lesssim32.4\,\rm MHz^{-1}$)} if measured by the uGMRT and $t_{\rm int}=500\,\rm hr$ (SKA1-low and $t_{\rm int}=50\,\rm hr$) if $\langle x_{\rm HI}\rangle=0.25$ and $\mathrm{log}_{10}f_{\rm X}=-2$, and hence should be detectable. In our modelling this corresponds to $T_{\rm HI}\approx30\,\rm K$.
    \item Making use of MCMC methods and exploring a wide range of parameter space, we have shown that a measurement of 10 $P_{21}$ of $22.1\,\rm MHz$ bandwidth can be used to constrain both the thermal (i.e. $\mathrm{log}_{10}f_{\rm X}$) and ionization (i.e. $\langle x_{\rm HI}\rangle$) state of the IGM at the same time. While the uncertainty arising from the telescope noise plays an important role in our parameter inference, the sample variance is the dominating effect. Therefore, observations by the SKA1-low over $50\,\rm hr$ \MNRASreply{per source} do not significantly improve the constraining power of the $P_{21}$ relatively to the uGMRT \MNRASreply{$500\,\rm hr$ per source} observations. Despite this, the SKA1-low should still have a better chance of detecting the signal.
    \item On the other hand, we suggest that a null-detection of the $P_{21}$ may provide interesting constraints on the properties of the high-$z$ IGM by disfavouring parts of the parameter space. Fig.~\ref{fig:multi_param_estimation_uGMRT_500h} and~\ref{fig:multi_param_estimation_SKA1-low}, \MNRASreply{particularly the solid black curve,} indicate that a null-detection of the $P_{21}$ may be utilized to: a) put a tighter upper limits on the $\langle x_{\rm HI}\rangle$ than are currently available from the \Lya~forest observations and b) put a lower limit on $\mathrm{log}_{10}f_{\rm X}$ which is not yet constrained well. For example, a null-detection from the observation of 10 RLQSO 21-cm forest power spectra over $500\,\rm hr$ \MNRASreply{each by the uGMRT can disfavour $T_{\rm HI}\lesssim54\,\rm K$ at $\langle x_{\rm HI}\rangle\approx0.25$, $T_{\rm HI}\lesssim80\,\rm K$ at $\langle x_{\rm HI}\rangle\approx0.50$ and $T_{\rm HI}\lesssim104\,\rm K$ at $\langle x_{\rm HI}\rangle\approx0.90$.} With $50\,\rm hr$ of observations by the SKA1-low, one can disfavour somewhat larger parameter space than in the previous case, \MNRASreply{particularly $T_{\rm HI}\lesssim94\,\rm K$ at $\langle x_{\rm HI}\rangle\approx0.25$, $T_{\rm HI}\lesssim141\,\rm K$ at $\langle x_{\rm HI}\rangle\approx0.50$ and $T_{\rm HI}\lesssim181\,\rm K$ at $\langle x_{\rm HI}\rangle\approx0.90$.} 
    \item Furthermore, even with $t_{\rm int}=50\,\rm hr$ of uGMRT time on 10 RLQSO, one can already start disfavouring models with a relatively \MNRASreply{high neutral hydrogen fraction, e.g. $\langle x_{\rm HI}\rangle\gtrsim0.18$ for a very cold IGM, and a very low X-ray preheating of the IGM. Specifically, if such observation would result in a null-detection, a parameter space of $T_{\rm HI}\lesssim17\,\rm K$ at $\langle x_{\rm HI}\rangle\approx0.25$, $T_{\rm HI}\lesssim26\,\rm K$ at $\langle x_{\rm HI}\rangle\approx0.50$ and $T_{\rm HI}\lesssim34\,\rm K$ at $\langle x_{\rm HI}\rangle\approx0.90$} can be disfavoured. This would be an important improvement in our understanding of the high-$z$ IGM for a modest observational time requirement. 
\end{itemize}

Note that we do not consider the contribution of minihaloes to the 21-cm forest absorption, however their significance in this matter is yet to be determined. Solving this is hindered by challenges of resolving such small scale structures in cosmological simulations and the fact that their contribution may be greatly reduced by feedback processes \citep{Meiksin_2011,Park_2016,Nakatani_2020,Chan_2024}. In addition, even structures within the minihaloes, subhaloes, may enhance the 21-cm forest absorption \citep{Kadota_2023}, but these are subject to tidal and/or ram pressure stripping \citep{Naruse_2024}.

Given that the 21-cm forest is expected to be non-Gaussian, using a multivariate Gaussian likelihood as in Eq.~\ref{eq:likelihood_covar} may be not an optimal method to extract information from the $P_{21}$\citep{Wolfson_2023}. As alternatives one can utilize a different type of the likelihood or a likelihood-free techniques such as in \citet{Sun_2024}.

Furthermore, while we implement various effects that can affect the observability of the 21-cm forest signal, such as the intrinsic radio spectrum of the background quasar and instrumental features, there are other effects that need to be considered. For example, the radio frequency interference (RFI), can introduce both the narrow-band and broad-band features in the radio spectrum of the background source. These can be partially mitigated (e.g. parts of the 21-cm spectrum affected by the narrow-band RFI features can be filtered out), however a detailed study of these effects is beyond the scope of this manuscript. Another instrumental effect which is omitted in this study is the contamination caused by foreground objects in the observed field via mode-mixing between parallel and transverse $k$-modes \citep{Thyagaragan_2020}.

We note that there were attempts of detecting the 21-cm forest absorption in the past. For example, \citet{Carilli_2007} measured a radio spectrum at frequencies where one would expect the 21-cm forest features of two radio-loud AGNs at $z=5.11$ and $z=5.2$ with the GMRT but they report no detection. This is not surprising because at these redshifts the IGM is already reionized and the signal is completely suppressed. Furthermore, their spectra have bandwidths of $\Delta\nu<5\,\rm MHz$. This means that they probe the IGM close enough to the host source that its radiation may heat up the gas even further \citep{Soltinsky_2023}. In this manuscript we have modelled significantly longer spectra, particularly with $\Delta\nu=22.1\,\rm MHz$. However, as one can see in Fig.~\ref{fig:1DPS_noise}, the $kP_{21}$ from the signal does not fall faster than the noise limit at \MNRASreply{$k\leq3.2\,\rm MHz^{-1}$}. From this we \MNRASreply{infer} that longer spectra (lower $k$) may improve our Bayesian parameter inference. Note that for longer spectra the effect of redshift evolution along the LOS would be non-negligible. To capture this effect lightcone simulations would be more suitable than the simulations utilized here. The uGMRT band-2 spans to $250\,\rm MHz$ \citep{Gupta_2017_uGMRT} which could provide at least twice as long spectra for sources at $z\approx6$ than assumed in our work. Unfortunately, at high enough frequencies we would start probing redshifts at which the reionization is already completed. This issue can be alleviated by observing higher redshift RLQSO. Note that there are now 3 RLQSO discovered at $z>6.8$ \citep{Banados_2021,Banados_2024,Endsley_2023}. We leave this for the future work.

We believe that the forecasts in this study will aid in similar observational efforts to finally acquire the first detection of the 21-cm forest signal at high redshift. On the other hand, we suggest that a null-detection of the signal can be translated into lower limit on the $f_{\rm X}$ and the upper limit on the $\langle x_{\rm HI}\rangle$. The former one can be instead changed to the lower limit on temperature of the neutral IGM at $z\gtrsim6$, which so far cannot be traced by other observables. Therefore, both outcomes of such observations would be very useful for the high-$z$ IGM studies.

Despite the caveats and challenges listed above, the prospects of achieving a statistical detection of the 21-cm forest are improving on theoretical (late-end reionization), observational (rising number of identified radio-loud quasars at high redshift) and instrumentational (improved sensitivity of the uGMRT, forthcoming construction of the SKA) fronts. These improvements in the field in combination with the potential of constraining the IGM properties at $z\geq6$ increase the importance of the 21-cm forest signal.

\section*{Acknowledgements}

We thank Arvind Balasubramanian, Silvia Belladitta, Abhirup Datta, Frederick Davies, Dominika Ďurovčíková, \MNRASreply{Keith Grainge,} Nissim Kanekar, Nabendu Kumar Khan, Shikhar Mittal and Jose Oñorbe for helpful discussions. \MNRASreply{Furthermore, we thank the anonymous referee for constructive comments which improved this manuscript.} We appreciate the feedback during the Astronomical Society of India meeting 2024 (\href{https://astron-soc.in/asi2024/}{https://astron-soc.in/asi2024/}), Cosmology in the Alps 2024 (\href{https://indico.skatelescope.org/event/1098/}{https://indico.skatelescope.org/event/1098/}), Cosmic Dawn at High Latitudes (\href{https://indico.fysik.su.se/event/8499/}{https://indico.fysik.su.se/event/8499/}), The First Gigayear(s) (\href{https://noirlab.edu/science/events/websites/first-gigayears-2024}{https://noirlab.edu/science/events/websites/first-gigayears-2024}) and \MNRASreply{SKA Cosmology SWG Meeting 2024 (\href{https://indico.ict.inaf.it/event/3054/}{https://indico.ict.inaf.it/event/3054/})} conferences which also shaped this manuscript. TŠ and GK gratefully acknowledge support by the Max Planck Society via a partner group grant. TŠ and GK are also partly supported by the Department of Atomic Energy (Government of India) research project with Project Identification Number RTI 4002. TŠ also expresses his gratutide for being supported by the Istituto Nazionale di Astrofisica - Osservatorio Astronomico di Trieste (INAF-OATs) under the Theory grant - 'Cosmological Investigation of the Cosmic Web' (C93C23006820005). We note that this work is heavily based on the \textsc{21cmfast} simulations \citep{Mesinger_2011_21CMFAST,Murray_2020}. The authors also thank the developers of \textsc{astropy} \citep{Robitaille_2013}, \textsc{corner} \citep{ForemanMackey_2016_corner}, \textsc{emcee} \citep{Foreman-Mackey_2013_emcee}, \textsc{matplotlib} \citep{Hunter_2007},  \textsc{numpy} \citep{Harris_2020} and \textsc{scipy} \citep{Virtanen_2020} packages which were implemented in the analyses in this work. This work has been performed as part of the DAE-STFC Technology and Skills Programme ‘Building Indo-UK collaborations towards the Square Kilometre Array’.

\section*{Data Availability}

The codes used for our analyses and for generating data and plots are publicly available at \href{https://github.com/tomassoltinsky/21cm-forest_1DPS}{https://github.com/tomassoltinsky/21cm-forest\textunderscore 1DPS}. All data used in this study can be requested from the first author.



\bibliographystyle{mnras}
\bibliography{example} 




\appendix

\MNRASreplyreply{\section{Convergence test of the IGM simulation for the 21-cm forest power spectrum}\label{app:convergence_test}}

\begin{figure}
     \begin{minipage}{1\columnwidth}
 	  \centering
 	  \includegraphics[width=\linewidth]{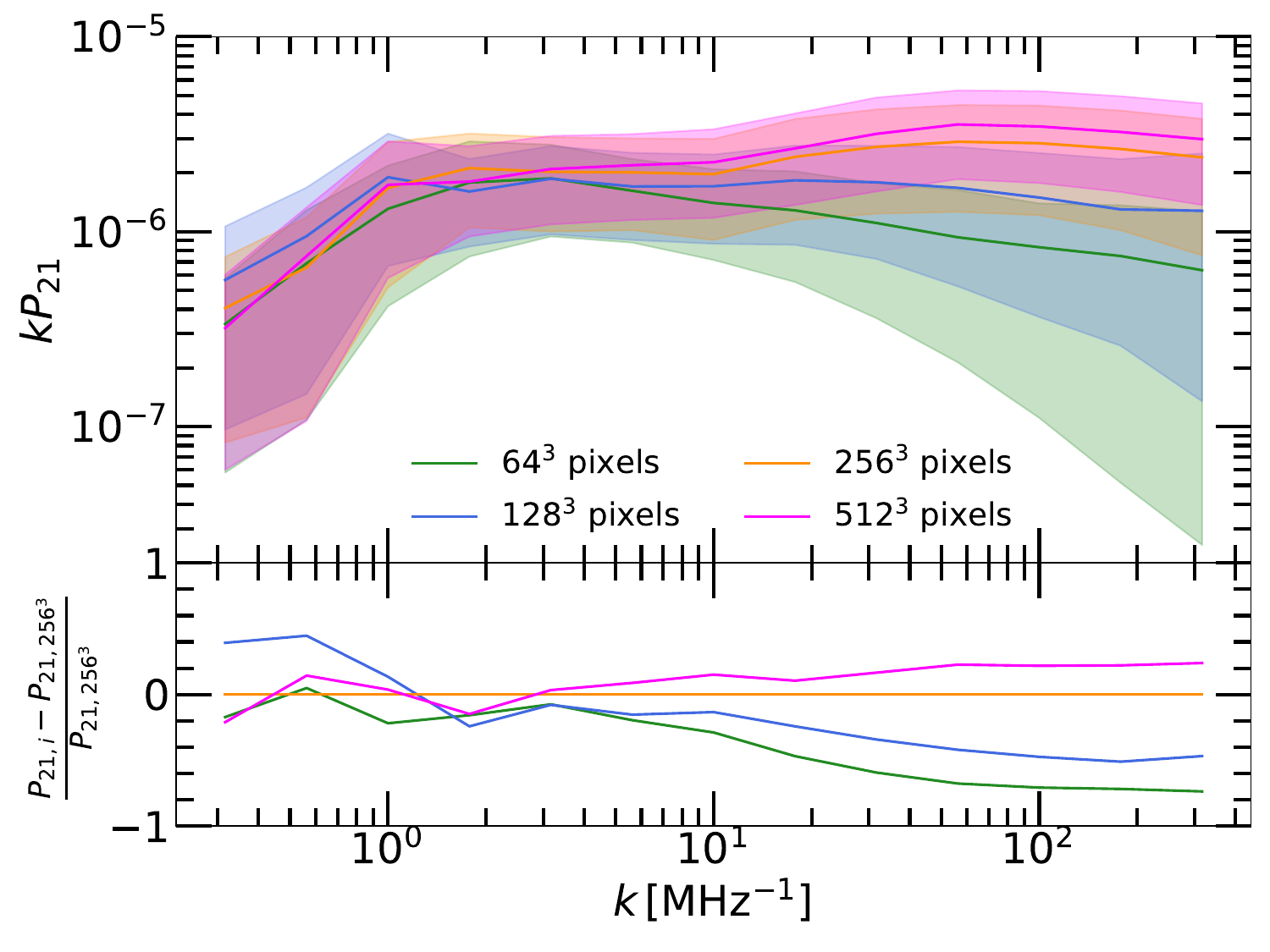}
	\end{minipage}
    \vspace{-0.3cm}
    \caption{\MNRASreplyreply{\textit{Top}: The convergence test of the \textsc{21cmfast} simulations. Particularly, we show the mean 21-cm forest 1D power spectrum, $kP_{21}$ (solid curves) and $68\%$ scatter (shaded regions with corresponding colours) calculated from 1000 mock spectra for the fiducial model of the IGM (i.e. $\mathrm{log}_{10}f_{\rm X}=-2$ and $\langle x_{\rm HI}\rangle=0.25$). We compare simulations of $50\,\rm cMpc$ volume and vary the number of pixels to $64^3$ (green), $128^3$ (blue), $256^3$ (orange, used in the paper) and $512^3$ (fuchsia). \textit{Bottom}: The fractional residuals computed relatively to the simulation used in the paper (i.e. $256^3$ pixels, orange curve).}}
    \label{fig:restest}
\end{figure}

\MNRASreplyreply{We have performed a convergence test to investigate the effect of the simulation resolution on the 1D 21-cm forest power spectrum. This is presented in Fig.~\ref{fig:restest} where we change the resolution of the simulation by fixing the simulated volume at $50\,\rm cMpc$ and varying the number of pixels to $64^3$ (green), $128^3$ (blue), $256^3$ (orange, used in this study) and $512^3$ (fuchsia). The top panel shows the $kP_{21}$ with the same $k$-binning as used in the paper while the bottom panel shows the residuals relative to the simulation used in the paper defined as $(P_{21,i}-P_{21,256^3})/P_{21,256^3}$, where $P_{21,i}$ is the 1D 21-cm forest power spectrum for a simulation with the number of pixels equal to $i$. The residuals indicate that our fiducial simulation underestimates the signal by up to $4\%$ at $k\approx1\,\rm MHz^{-1}$ and up to $9\%$ at $k\approx6\,\rm MHz^{-1}$ compared to the simulation with the highest resolution considered (i.e. $512^3$ pixels, fuchsia). However, note that the discrepancy between the $512^3$ and $256^3$ simulations is significantly smaller than the uncertainty due to the sample variance (indicated by the shaded regions).}

\MNRASreplyreply{We also emphasize that this study focuses on taking the advantage of the advancements in the field which improve the prospects of detecting the 21-cm forest signal and exploration of what can be implied from the (null-)detection of this signal rather than the detailed modelling of the physics in the simulation. A more accurate modelling, such as the use of hydrodynamics/radiative transfer simulations including \textsc{Nyx} \citep{Onorbe_2019}, \textsc{CoDa III} \citep{Lewis_2022_CoDaIII}, \textsc{Sherwood-relics} \citep{Puchwein_2023} and \textsc{THESAN} \citep{Garaldi_2024_THESAN} to name a few. Unfortunately, even high resolution simulations with similar volumes as considered in our study do not resolve minihaloes due to large dynamic range they are required to capture. This is one of the big challenges currently in the numerical cosmology community. Furthermore, such simulations would be impossible to run for as large parameter space as desired for our study due to computational costs. Hence, we decided to utilize the \textsc{21cmfast} instead. Running the \textsc{21cmfast} simulation with finer resolution would be computationally too expensive too and would not improve the analysis very much. In spite of this, we plan to use better simulations (more detailed physics, wider dynamic range) for such analysis in the future.\\}


\section{The effect of redshift space distortions on the 21-cm forest power spectrum}\label{app:vpec}

\begin{figure}
     \begin{minipage}{1\columnwidth}
 	  \centering
 	  \includegraphics[width=\linewidth]{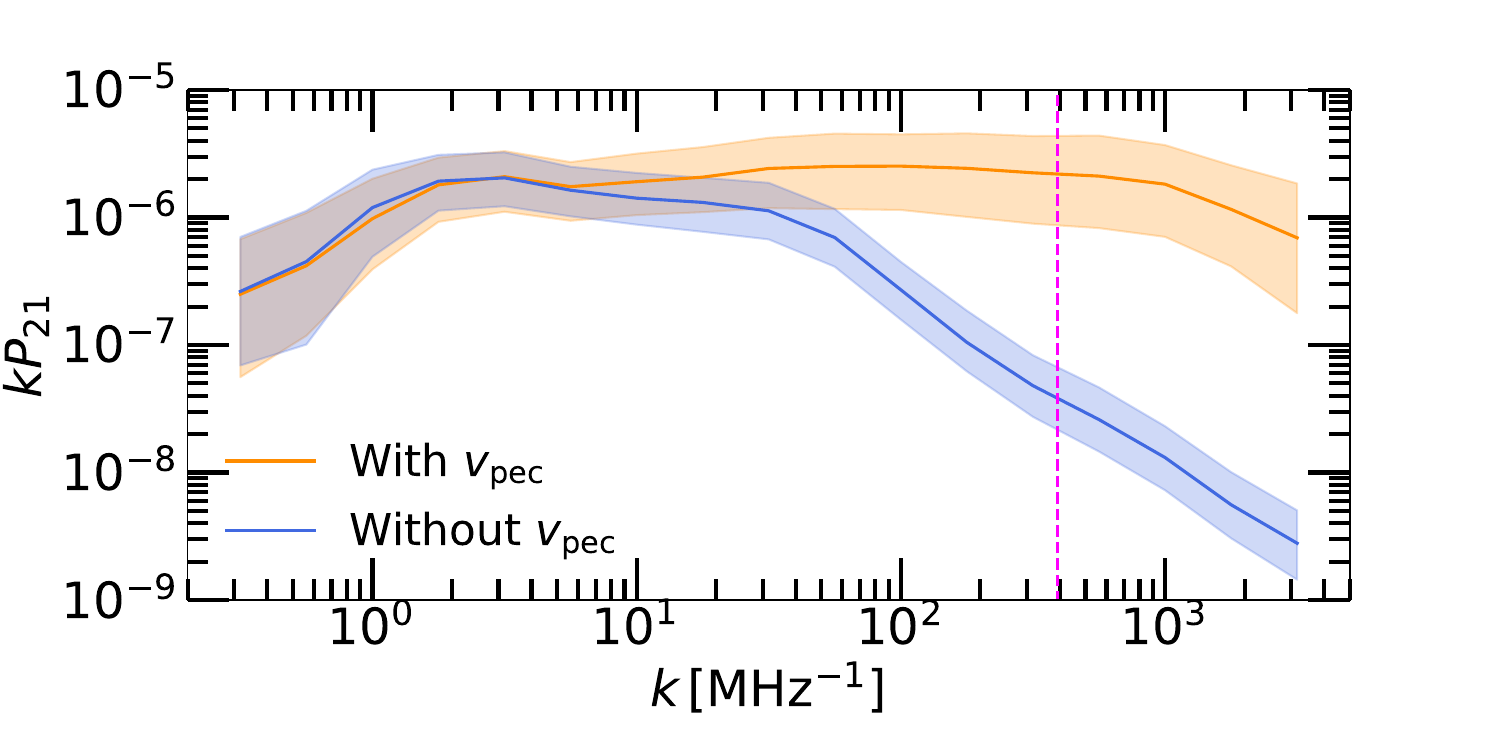}
	\end{minipage}
	\vspace{-0.3cm}
    \caption{21-cm forest 1D power spectrum including (orange curve) and neglecting (blue curve) peculiar velocities. The solid lines represent median values and the shaded regions mark 68 per cent range from 2000 mock spectra. The vertical dashed fuchsia line marks the highest $k$ that is accessible by a telescope given a spectral resolution $\Delta\nu=8\,\rm kHz$.}
    \label{fig:1DPS_vpec}
\end{figure}

The redshift space distortions can affect the 21-cm forest signal by both shifting the absorption lines relative to the location of the absorbers from which they arise in redshift space and significantly boosting the optical depth of the 21-cm line absorption features \citep{Semelin_2016,Soltinsky_2021}. In this section we test if the redshift space distortions have a substantial effect on the 21-cm forest power spectrum or if the effect of peculiar velocities of the gas can be neglected. In Fig.~\ref{fig:1DPS_vpec} we show the $P_{21}$ for the case in which we neglect the peculiar velocity of the gas (i.e. $v_{\rm pec}=0$, blue) and if we include it in our computation (orange). The effect of redshift space distortions, if any, is expected to be more pronounced at small scales due to shifts of absorption features in the redshift space. Therefore, we deliberately do not smooth our synthetic 21-cm forest spectra such that we access the largest $k$ values. One can clearly see in Fig.~\ref{fig:1DPS_vpec} that the redshift space distortions increase the $P_{21}$ at \MNRASreply{$k\gtrsim20\,\rm MHz^{-1}$} as they force the absorption features to cluster more. This might lead to misinterpretation of the observations, especially the dark matter constraints as described in \citet{Shao_2023}, who neglect the redshift space distortions, which come from the large $k$ part of the $P_{21}$. In addition, the discrepancy between the $v_{\rm pec}=0$ and $v_{\rm pec}\neq0$ case extends to $k=393\,\rm MHz^{-1}$, corresponding to the spectral resolution of $\Delta\nu=8\,\rm kHz^{-1}$ assumed in this study (vertical dashed fuchsia line). Hence, neglecting $v_{\rm pec}$ would affect the analyses of observations like we describe in this work too.


\MNRASreply{\section{Signal-to-noise ratio of the 21-cm forest}\label{app:SNR}}

\begin{figure}
     \begin{minipage}{1\columnwidth}
 	  \centering
 	  \includegraphics[width=\linewidth]{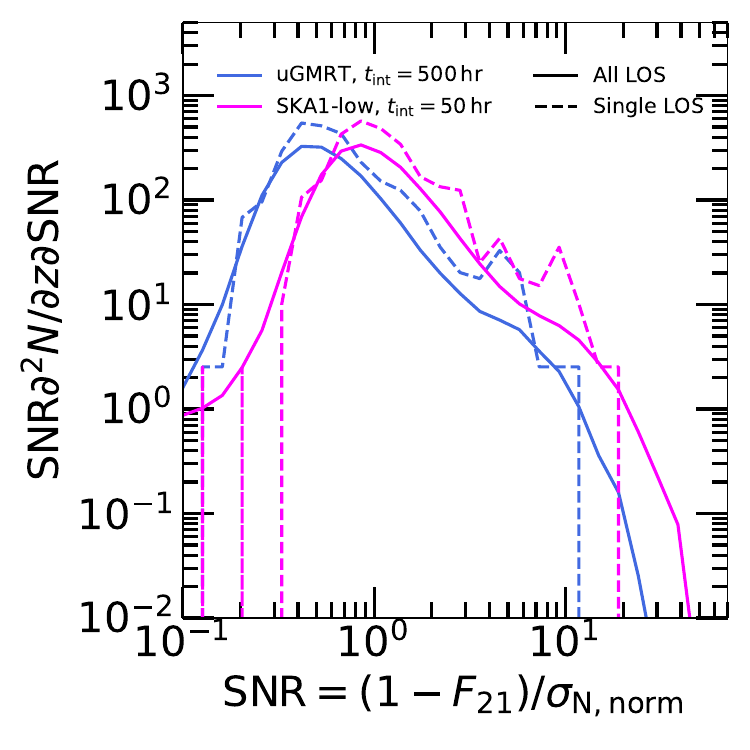}
	\end{minipage}
	\vspace{-0.3cm}
    \caption{\MNRASreply{The differential number density of the SNR of the simulated 21-cm forest spectrum per frequency channel. The blue curves correspond to the noise assuming an observation by the uGMRT over $500\,\rm hr$ and the fuchsia curves correspond to the observation by the SKA1-low over $50\,\rm hr$. In both cases we consider that the radio telescopes target a $z=6$ quasar with the spectrum defined by $S_{147}=64.2\,\rm mJy$ and $\alpha_{\rm R}=-0.44$ in the IGM with $\mathrm{log}_{10}f_{\rm X}=-2$ and $\langle x_{\rm HI}\rangle=0.25$. The solid curves show distribution for 1000 synthetic LOS while the dashed curves are for a single LOS shown in Fig.~\ref{fig:noisyspectrum}.}}
    \label{fig:SNR_dist}
\end{figure}

\MNRASreply{To quantify the difficulty of a direct detection of individual absorption features of the 21-cm forest we compute the signal-to-noise ratio (SNR) for each frequency channel in the synthetic spectra. We define $\mathrm{SNR}(\nu)=(1-F_{21}(\nu))/\sigma_{\rm N,\ norm}(\nu)=(1-e^{-\tau_{21}(\nu)})/\sigma_{\rm N,\ norm}(\nu)$, where $\sigma_{\rm N,\ norm}$ is the noise rms computed from Eq.~\ref{eq:noise} normalized with the intrinsic spectrum of the background radio source. Note that the $\sigma_{\rm N,\ norm}(\nu)$ depends on the frequency $\nu$ given the dependency of the telescope sensitivity on the $\nu$ and the normalization by the background source continuum.}

\MNRASreply{In Fig.~\ref{fig:SNR_dist} we show the differential number density of the SNR per frequency channel for the IGM model of $\mathrm{log}_{10}f_{\rm X}=-2$ and $\langle x_{\rm HI}\rangle=0.25$. Firstly, we consider a single LOS, specifically the one shown in Fig.~\ref{fig:noisyspectrum} (dashed curves). We see that the highest SNR reached using the observational setup of the uGMRT, $t_{\rm int}=500\,\rm hr$ is $11.9$ and for the SKA1-low, $t_{\rm int}=50\,\rm hr$ it is $19.0$. While the absorption features with $\mathrm{SNR}\approx10$ could be detectable, frequency channels with such SNR are $\sim20$ times less abundant than the ones at the peak of the distribution which are at $\mathrm{SNR}\approx0.4$ and $0.9$ for the uGMRT, $t_{\rm int}=500\,\rm hr$ and the SKA1-low, $t_{\rm int}=50\,\rm hr$, respectively. There is a similar trend if we consider all 1000 simulated LOS in the same IGM model (solid curves). There are some LOS that reach SNR of $36.5$ if observed by the uGMRT over $500\,\rm hr$ and $63.3$ if observed by the SKA1-low over $50\,\rm hr$. Unfortunately, these are still very rare. Hence, the direct detection of the individual absorption features of the 21-cm forest seems very difficult.}


\bsp	
\label{lastpage}
\end{document}